\documentclass[useAMS,usedcolumn,usegraphicx,usenatbib]{mn2e}

\usepackage{times}
\usepackage{amsmath}                  % mathematischer Zeichensatz
\usepackage{amssymb}                  % mathematische Symbole
\usepackage{graphicx}
\usepackage{units}
\usepackage{dcolumn}
\usepackage{longtable}
\usepackage{lscape}
\usepackage{booktabs}
\usepackage{subfigure}
\usepackage{natbib}
\usepackage{aas_macros}
\usepackage[draft,pdftex,pdfpagemode={UseOutlines},bookmarks,bookmarksopen,colorlinks,linkcolor={blue},citecolor={green},urlcolor={red}]{hyperref}
\let\orgautoref\autoref
\providecommand{\Autoref}
        {\def\equationautorefname{Equation}%
         \def\figureautorefname{Fig.}%
         \def\subfigureautorefname{Fig.}%
         \def\partautorefname{Part}%
         \def\chapterautorefname{Chapter}%
         \def\sectionautorefname{Section}%
         \def\subsectionautorefname{Section}%
         \def\subsubsectionautorefname{Section}%
         \def\Itemautorefname{Item}%
         \def\tableautorefname{Table}%
         \def\lstlistingautorefname{Listing}%
         \orgautoref}
\renewcommand{\autoref}
        {\def\equationautorefname{equation}%
         \def\figureautorefname{Fig.}%
         \def\subfigureautorefname{Fig.}%
         \def\partautorefname{part}%
         \def\chapterautorefname{chapter}%
         \def\sectionautorefname{section}%
         \def\subsectionautorefname{section}%
         \def\subsubsectionautorefname{section}%
         \def\appendixautorefname{appendix}%
         \def\Itemautorefname{item}%
         \def\tableautorefname{Table}%
         \def\lstlistingautorefname{Listing}%
         \orgautoref}
\providecommand{\autorefs}
        {\def\equationautorefname{equations}%
         \def\figureautorefname{Figs.}%
         \def\subfigureautorefname{Figs.}%
         \def\partautorefname{parts}%
         \def\chapterautorefname{chapters}%
         \def\sectionautorefname{sections}%
         \def\subsectionautorefname{sections}%
         \def\subsubsectionautorefname{sections}%
         \def\Itemautorefname{items}%
         \def\tableautorefname{Tables}%
         \def\lstlistingautorefname{Listings}%
         \orgautoref}

\makeatletter
\newcolumntype{d}[1]{>{\DC@{,}{{.}}{#1}}c<{\DC@end}}
\newcolumntype{o}[1]{>{\DC@{+}{\pm}{#1}}c<{\DC@end}}
\makeatother
\makeatletter
\newcolumntype{f}[1]{>{\DC@{p}{\ldots}{#1}}c<{\DC@end}}
\makeatother

\title[A catalogue of young runaway stars]{A catalogue of young runaway Hipparcos stars within 3kpc from the Sun}
\author[N. Tetzlaff, R. Neuh\"auser and M. M. Hohle]{N. Tetzlaff$^{1}$\thanks{E-mail:
nina@astro.uni-jena.de}, R. Neuh\"auser$^{1}$ and M. M. Hohle$^{1,2}$\\
$^{1}$Astrophysikalisches Institut und Universit\"ats-Sternwarte Jena, Schillerg\"asschen 2-3, 07745 Jena, Germany\\
$^{2}$Max-Planck-Institut f\"ur extraterrestrische Physik, Giessenbachstra{\ss}e, 85741 Garching, Germany}

\begin{document}

\date{Accepted. Received; in original form}

\pagerange{\pageref{firstpage}--\pageref{lastpage}} \pubyear{2010}

\maketitle

\label{firstpage}

\begin{abstract} 

Traditionally runaway stars are O and B type stars with large peculiar velocities. We want to extend this definition to young stars (up to $\approx\unit[50]{Myr}$) of any spectral type and identify those present in the Hipparcos catalogue applying different selection criteria such as peculiar space velocities or peculiar one-dimensional velocities.\\
Runaway stars are important to study the evolution of multiple star systems or star clusters as well as to identify origins of neutron stars.\\
We compile distances, proper motions, spectral types, luminosity classes, $V$ magnitudes and $B-V$ colours and utilise evolutionary models from different authors to obtain star ages and study a sample of 7663 young Hipparcos stars within $\unit[3]{kpc}$ from the Sun. Radial velocities are obtained from the literature.\\
We investigate the distributions of the peculiar spatial velocity, the peculiar radial velocity as well as the peculiar tangential velocity and its one-dimensional components and obtain runaway star probabilities for each star in the sample. In addition, we look for stars that are situated outside any OB association or OB cluster and the Galactic plane as well as stars of which the velocity vector points away from the median velocity vector of neighbouring stars or the surrounding local OB association/ cluster although the absolute velocity might be small. \\
We find a total of 2547 runaway star candidates (with a contamination of normal Population I stars of 20 per cent at most). Thus, after subtraction of those 20 per cent, the runaway frequency among young stars is about 27 per cent.\\
We compile a catalogue of runaway stars which will be available via VizieR.

\end{abstract}


%%%%%%%%%%%%%%%%%%%%%%%%%%%%%%%%%%%%%%%%%%%%%%%%%%%%%%%%%%%%%%%%%%%%%%%%%%%%%%%%%%%%%%%%%%%%%%%%%%%%%%%%%%%%%%%%%%%%%%%%%%%%%%%%%%%%%%%%%%%%%%%%%

\section{Introduction}\label{sec:intro}

Almost fifty years ago, \citet{1961BAN....15..265B} found that many O and B type stars show large peculiar space velocities ($>\unit[40]{km/s}$). For that reason they were assigned the term ``runaway'' stars. Many studies concerning O and B type runaway stars have been published since then covering different selection methods \citep[e.g.][]{1965IzKry..34..193V,1974RMxAA...1..211C,1986ApJS...61..419G,1991AJ....102..333S,1998AA...331..949M}.\\
Two theories on the formation of runaway stars are accepted:
\begin{enumerate}
\item The binary supernova (SN) scenario (BSS) \citep{1961BAN....15..265B} is related to the formation of the high velocity neutron stars: The runaway and the neutron star are the products of a SN within a binary system. The velocity of the former secondary (the runaway star) is comparable to its original orbital velocity. \newline Runaway stars formed in the BSS share typical properties such as enhanced helium abundance and a high rotational velocity due to mass and angular momentum transfer during binary evolution. The kinematic age of the runaway star is smaller than the age of its parent association.
\item In the dynamical ejection scenario (DES) \citep{1967BOTT....4...86P} stars are ejected from young, dense clusters via gravitational interactions between the stars. \newline The (kinematic) age of a DES runaway star should be comparable to the age of its parent association since gravitational interactions occur most efficiently soon after formation.
\end{enumerate}
Which scenario is dominating is still under debate; however, both are certainly taking place (one example for each identified by \citealt{2001AA...365...49H}: BSS -- PSR B1929+10/ $\zeta$ Oph; DES -- AE Aur/ $\mu$ Col/ $\iota$ Ori).\\

The selection criteria for runaway stars of previous studies were either based on spatial velocities \citep[e.g.][]{1961BAN....15..265B}, tangential velocities \citep[e.g.][]{1998AA...331..949M} or radial velocities \citep[e.g.][]{1974RMxAA...1..211C} alone. According to \citet{1979ApJ...232..520S} the velocity distribution of early type stars can be explained with the existence of two different velocity groups of stars: a low velocity group containing normal Population I stars and a high velocity group containing the runaway stars (see, e.g., \autoref{fig:2Maxwell}). Since both groups obey a Maxwellian velocity distribution also runaway stars with relatively low velocities exist. We want to combine previous methods in order not to miss an important star because the radial velocity may be unknown (hence no spatial velocity) or its tangential or radial velocity component may be significantly larger than the other component (hence one would miss it in one direction). Moreover, we also want to identify the lower velocity runaway stars by searching for stars that were ejected slowly from their parent cluster. Furthermore, we want to use the term ``runaway'' star not only for O and B type runaway stars but all young (up to $\approx\unit[50]{Myr}$) runaway stars to account for the possibility of less massive companions of massive stars (which soon after formation explode in a SN) and low-mass stars in young dense clusters which also may be ejected due to gravitational interactions.\\
In \autoref{sec:sample} we describe the selection procedure of our sample of young stars. In \autoref{sec:runaways} we apply different identification methods for runaway stars and assign runaway probabilities to each star. A catalogue containing our results will be available via VizieR soon after publication. \\
We give a summary and draw our conclusions in \autoref{sec:conclusions}.

%__________________________________________________________________

\section{The sample of young stars}\label{sec:sample}

We start with all stars from the Hipparcos catalogue \citep{1997A&A...323L..49P}, 118218 in total, and collect spectral types as well as $V$ magnitudes and $B-V$ colours from the catalogue. According to the errata file provided with the Hipparcos catalogue, we correct erroneous spectral types. From the new Hipparcos reduction \citep{2007AA...474..653V} we obtain parallaxes ($\pi$) and proper motions ($\mu_\alpha^\ast=\mu_\alpha\cos\delta$, $\mu_\delta$). We restrict our star sample to lie within $\unit[3]{kpc}$ from the Sun ($\pi-\sigma_\pi\geq\unit[1/3]{mas}$ with $\sigma_\pi$ being the $1\sigma$ error on $\pi$).\footnote{Since we are working with individual stars we do not correct for statistical biases \citep[][]{1996MNRAS.281..211S}. In any case, for the stars in our sample the Smith-Eichhorn corrected parallaxes do not differ significantly from the measured ones.} Furthermore, we remove stars in the regions of the Large and Small Magellanic Clouds having accidentally $\pi-\sigma_\pi\geq\unit[1/3]{mas}$ \citep[cf.][]{2010AN....331..349H}. This gives us an initial set of 103217 stars.\\
In the cases where the Hipparcos catalogue does not provide sufficient spectral types, we take the spectral type from either the Simbad or VizieR databases \citep{2009yCat....102023S,2009yCat.1280....0K,2007yCat.3246....0A,2007AN....328..889K,2006ARep...50..733B,2003AJ....125..359W,1999mscf.book.....B,1999AAS..137..451G,1982ApJ...263..777G}. Missing $B-V$ colours were amended from different sources \citep[Simbad; available via VizieR:][]{2007AN....328..889K,2004AN....325..439K,2006ARep...50..733B,2005yCat.1297....0Z,2003AJ....125..359W,1994yCat.1197....0E}.\\
Having now collected parallaxes, spectral types as well as $V$ magnitudes and $B-V$ colours we calculate luminosities ($L$) and obtain effective temperatures ($T_{eff}$) from spectral types according to \citet{Schmidt-Kaler1982} and \citet{1995ApJS..101..117K}. The extinction $A_V$ is determined from the apparent $B-V$ colour and the spectral type.\footnote{For some stars with unknown luminosity class we assume luminosity class V. For the effective temperature $T_{eff}$ differences between different luminosity classes are modest and the error of the luminosity $L$ is mainly caused by the error on the parallax.} The initial sample contains 1721 stars also included in the list of \citet{2010AN....331..349H} who used additional colours to determine $L$. Thus, for those stars, we adopt their luminosities (without Smith-Eichhorn correction).\\
For having full kinematics, we obtain radial velocities from Simbad or the following VizieR catalogues: \citet{2008AJ....136..421Z}, \citet{2007AN....328..889K,2004AN....325..439K}, \citet{2006ARep...50..733B}, \citet{2006AstL...32..759G}, \citet{2006yCat.3249....0M}, \citet{2004yCat..73491069K}, \citet{2000AAS..142..217B}, \citet{1999AAS..137..451G}, \citet{1997yCat.3198....0R} and \citet{1967IAUS...30...57E}.\\

\subsection{Age and mass determination}\label{subsec:agemass}

Since we look for young stars, our first goal is to obtain star ages. For that reason we check our star sample for pre-main sequence stars and find 236 of them either in catalogues  \citep{2010A&A...509A..52C,2005AA...438..769D,2005AJ....129..856H,2005yCat..33260211W,2003yCat..21470305V,2000AA...356..541K,2000AAS..146..323N,2000AA...361.1143T,1999AA...352..574B,1997A&A...324L..33V,1995ApJS..101..117K,1995AA...297..391N,1994AAS..104..315T} and/ or fulfilling the lithium criterion from \citet{1997Sci...276.1363N} (for individual stars see also \citealt{2003PhDTh...Koenig,2002A&A...384..491C,2002A&A...395..877N,2002A&A...385..862S,2001MNRAS.321...57L,2001A&A...379..976M,2001ApJ...549L.233Z,2000A&A...363..239D,2000AJ....120.1006G,2000A&AS..142..275S,2000AA...361.1143T,2000AJ....120.1410T,2000ApJ...535..959Z,1999ApJ...512L..63W,1998PASP..110.1259G,1998A&A...330L..29N,1998AJ....116..396S,1997A&A...326..647F,1997A&A...326..221M,1997A&A...321..850T,1995MNRAS.273..559J,1989ApJS...71..895W}). \\
For the 236 pre-main sequence stars in the sample we used pre-main sequence evolutionary models from \citet{1994ApJS...90..467D,1997MmSAI..68..807D}, \citet{1996AAS..120...57B}, \citet{1998AA...337..403B}, \citet{1999ApJ...525..772P}, \citet{2000AA...358..593S}, \citet{2002ASPC..267..179M} and \citet{2008Ap&SS.316..173M}\footnote{http://www.astro.up.pt/corot/models/cesam} to estimate their masses and ages. \\
For main sequence and post-main sequence stars we used evolutionary models starting from the zero-age main sequence (ZAMS) from \citet{1992AAS...96..269S}, \citet{2004AA...424..919C}, \citet{2004ApJ...612..168P}\footnote{http://albione.oa-teramo.inaf.it/}, \citet{2008AA...484..815B,2009AA...508..355B} and \citet{2008Ap&SS.316..173M}. For stars on or above the ZAMS all models yield masses and ages based on luminosity and effective temperature. Stars which lie below the model ZAMS (mainly due to large uncertainties in the parallax, thus in the luminosity) are shifted towards the ZAMS, hence will be treated as ZAMS stars. For early-type stars this has no effect on the age selection (see \autoref{eq:taulimit}) because they are younger than $\unit[50]{Myr}$ even if the actual position in the HRD would be above the ZAMS. For most mid-F to mid-G stars age determination is difficult and the error on the age is larger than the age value itself (expressing the time the star would evolve without moving in the HRD), thus they will not enter our list of young stars due to our age criterion (see \autoref{eq:taulimit}). Even later late-type stars (later than mid-G) are already older than $\unit[50]{Myr}$ on the ZAMS.\\
We used Solar metallicity for all 101628 stars in our sample with known magnitudes and spectral types; we obtained masses and ages from luminosities and temperatures. For stars with unknown luminosity class, we adopt luminosity class V\footnote{For the effective temperature $T_{eff}$ differences between different luminosity classes are modest and the error of the luminosity $L$ is mainly caused by the error on the parallax.}. For some stars where also the spectral type is uncertain to $\pm5$ sub-types, we calculate masses and ages for all spectral sub-types (e.g. for a G star: G0 to G9) to derive the mean and the standard deviation of mass and age.\\
We define a star to be ``young'' if its age is $\leq\unit[50]{Myr}$. This limit is set for the following 
%reasons arising from the two formation scenarios of runaway stars described in \autoref{sec:intro}. 
reason arising from the BSS formation scenario of runaway stars (\autoref{sec:intro}):
%\begin{enumerate}
%	\item BSS: \newline 
						It is desirable to identify the (now isolated) neutron star which was formed in the SN that released the runaway star. This would also yield the runaway's origin. For neutron stars the radial velocity is unknown, thus must be treated with a probability distribution \citep{2010MNRAS.402.2369T}. For that reason the error cone of the spatial motion is large and the position of a neutron star can be determined reliably only for a few million years, optimistically $\approx\unit[5]{Myr}$. This is the maximum runaway time (the kinematic age) of the runaway star (as well as the neutron star) such that the neutron star could be identified. The latest spectral type on the main-sequence for stars to explode in a SN and eventually become a neutron star is B3. These stars live about $\unit[35]{Myr}$ before they end their lives in SNe. We allow for an uncertainty of $\unit[10]{Myr}$ and find the maximum star age (not the kinematic age) of the BSS runaway for which the associated neutron star should still be identifiable to be $\approx\unit[50]{Myr}$.
Despite that, a larger age would mean a longer timespan to trace back the star to identify its origin (if it is a DES runaway). This would cause large error bars on the past position of the runaway star that would make it less reliable to find the origin.\\
However, star ages often suffer from large uncertainties due to large errors in distances and strong uncertainties in evolutionary models. That is why we choose the following criterion on the star age $\tau_\star$ (the median from all evolutionary models applied) for a star to be young within our definition:
\begin{equation}
	\mbox{both } \ \ \ \tau_\star + \sigma_{\tau,\star} \leq \unit[100]{Myr}\ \ \ \mbox{ and }\ \ \ \tau_\star - \sigma_{\tau,\star} \leq \unit[50]{Myr},\label{eq:taulimit}
\end{equation}
where $\sigma_{\tau,\star}$ is the median deviation of $\tau_\star$ for different models. With this criterion we allow for an error of $\tau_\star$ that is of the order of $\tau_\star$ itself but also exclude stars with accurately known ages above our limit of $\unit[50]{Myr}$. Unfortunately, ages of supergiants are rather uncertain and we miss many of them applying the criterion. For that reason we add all stars of luminosity classes I and II as well as stars of luminosity class III earlier than A0 because they could certainly be younger than $\unit[50]{Myr}$. Moreover, the classical definition of runaway stars by \citet{1961BAN....15..265B} includes stars up to B5 of luminosity classes IV and V. These are added as well. Finally, our sample contains 7663 young stars. Their Hipparcos numbers and common names as well as ages, masses and spectral types are listed in \autoref{tab:HIPid} (the full table will be available via the VizieR database soon after publication). Since 2250 stars in the table enter our list only due to their spectral type and luminosity class, we list only their spectral type and luminosity class as given in the literature.\\
\begin{table}
\centering
\caption{Ages $\tau_\star$ (in Myr), masses $M_\star$ (in solar masses) and spectral types for 7663 young stars (sorted by their HIP number). $\tau_\star$ and $M_\star$ are medians obtained from different evolutionary models (see \autoref{subsec:agemass}). For 2250 stars only the spectral type is given since models infer a larger age; however they are probably also young (as inferred from spectral type and luminosity class). Here the first five entries are shown, the full table will be available via VizieR soon after publication.}\label{tab:HIPid}
\begin{tabular}{r c d{1}<{\,\pm} >{\hspace{-0.1\tabcolsep}}d{3} d{1}<{\,\pm} >{\hspace{-0.1\tabcolsep}}d{3} c}
\toprule
HIP & other name & \multicolumn{2}{c}{$\tau_\star$ [Myr]}	&	\multicolumn{2}{c}{$M_\star$ [$\unit{M_\odot}$]} & SpT\\\midrule
32 & HD 224756 & 10,0 & 7,6 & 2,8 & 0,1 & B8 \\
85 & CD-25 16747 & 39,8 & 28,8 & 1,8 & 0,1 & A2 \\
106 & HD 224870 & \multicolumn{2}{c}{--} & \multicolumn{2}{c}{--} & G7II-III\\
124 & HD 224893 & 44,7 & 5,4 & 7,6 & 0,4 & F0III \\ 
135 & HD 224908 & 45,7 & 22,9 & 1,0 & 0,1 & G5 \\
\bottomrule
\end{tabular}
\end{table}

\subsection{Kinematics}\label{subsec:kinematicsyoungstars}

The solar motion with respect to a specific star sample, i.e. the Local Standard of Rest (LSR), depends upon the age of the stars in the sample \citep[e.g.][]{1981gask.book.....M}. For that reason, we derive the kinematic centre of the stars in our sample as follows. We calculate spatial velocity components of the 4195 stars with complete kinematic data (in a right-handed coordinate system with the $x$ axis pointing towards the Galactic centre and $y$ is positive in the direction of galactic rotation) corrected for Galactic differential rotation using Keplerian orbits:
\begin{eqnarray}
& \hat{U}  = & U - U_{rot},\notag\\
& \hat{V}  = & V - V_{rot} + V_{\odot,rot},\\
& \hat{W}  = & W,\notag
\end{eqnarray}
where $U$, $V$ and $W$ are the heliocentric velocity components (in the $x$, $y$ and $z$ direction, respectively) and $U_{rot}$ and $V_{rot}$ the components of the rotational velocity of the star moving around the Galactic centre. $V_{\odot,rot}=\unit[225]{km/s}$ is the rotational velocity of the Sun around the Galactic centre. To avoid significant contamination of high velocity stars, we exclude stars with $\hat{v}>\unit[50]{km/s}$ (that is approximately two times the median of $\hat{v}=\sqrt{\hat{U}^2+\hat{V}^2+\hat{W}^2}$).\\
Fitting a Gaussian to each velocity component, we find
\begin{equation}
\left(U_\odot,V_\odot,W_\odot\right) = \left(10{.}4\pm0{.}4,11{.}6\pm0{.}2,6{.}1\pm0{.}2\right)\,\mathrm{km/s}.\label{eq:LSR}
\end{equation}
The agreement of our LSR with the classical value of \citet{1965gast.conf...61D} [($9$,$11$,$6$)\,km/s] is remarkable. In comparison with the most recent value published by \citet{2009MNRAS.397.1286A} [($9{.}96\pm0{.}33$,$5{.}25\pm0{.}54$,$7{.}07\pm0{.}34$)\,km/s], the $V$ component differs significantly; however, \citet{2009MNRAS.397.1286A} obtained their results by examining the correlation between LSR and colour $B-V$, i.e. star age, and extrapolating the curves to zero velocity dispersion and also ignoring stars with $B-V\leq0$ \citep[see also][]{1998MNRAS.298..387D} because they are probably not yet mixed. Such stars are overabundant in our sample of young stars. From Figure 3 in \citet{1998MNRAS.298..387D} one may easily see that our result agrees well with their findings.\\
In the following section we will use \autoref{eq:LSR} to correct the velocities for Solar motion. The peculiar velocity is the velocity of a star corrected for Solar motion and Galactic rotation.

%______________________________________________________________

\section{Young runaway stars}\label{sec:runaways}

Classically, a star is assigned to be a runaway star if its peculiar space velocity $v_{pec}$ exceeds $\unit[40]{km/s}$ \citep{1961BAN....15..265B}. Later on, \citet{1965IzKry..34..193V} and \citet{1974RMxAA...1..211C} used radial velocities $v_r$ alone to investigate the population of runaway stars defining a more general definition of $\left|v_{r,pec}\right|>v_{crit,1D}$ with $v_{crit,1D}=3\sigma$, where $\sigma$ is the mean velocity dispersion in one dimension of the low velocity stars. Based on this definition \citet{1998AA...331..949M} investigated tangential velocities $v_t$ to identify runaway stars. We now want to combine and extend all these methods.

%%%%%%%%%%%%%%%%%%%%%%%%%%%%%%%%%%%%%%%%%%%%%%%%%%%%%%%%%%%%%%%%%%%%%%%%%%%%%%%%%%%%%%%

\subsection{Runaway stars identified from their peculiar space velocity}\label{subsec:3Dvel}

\begin{figure}
\centering
\includegraphics[width=0.45\textwidth, viewport= 40 200 535 610]{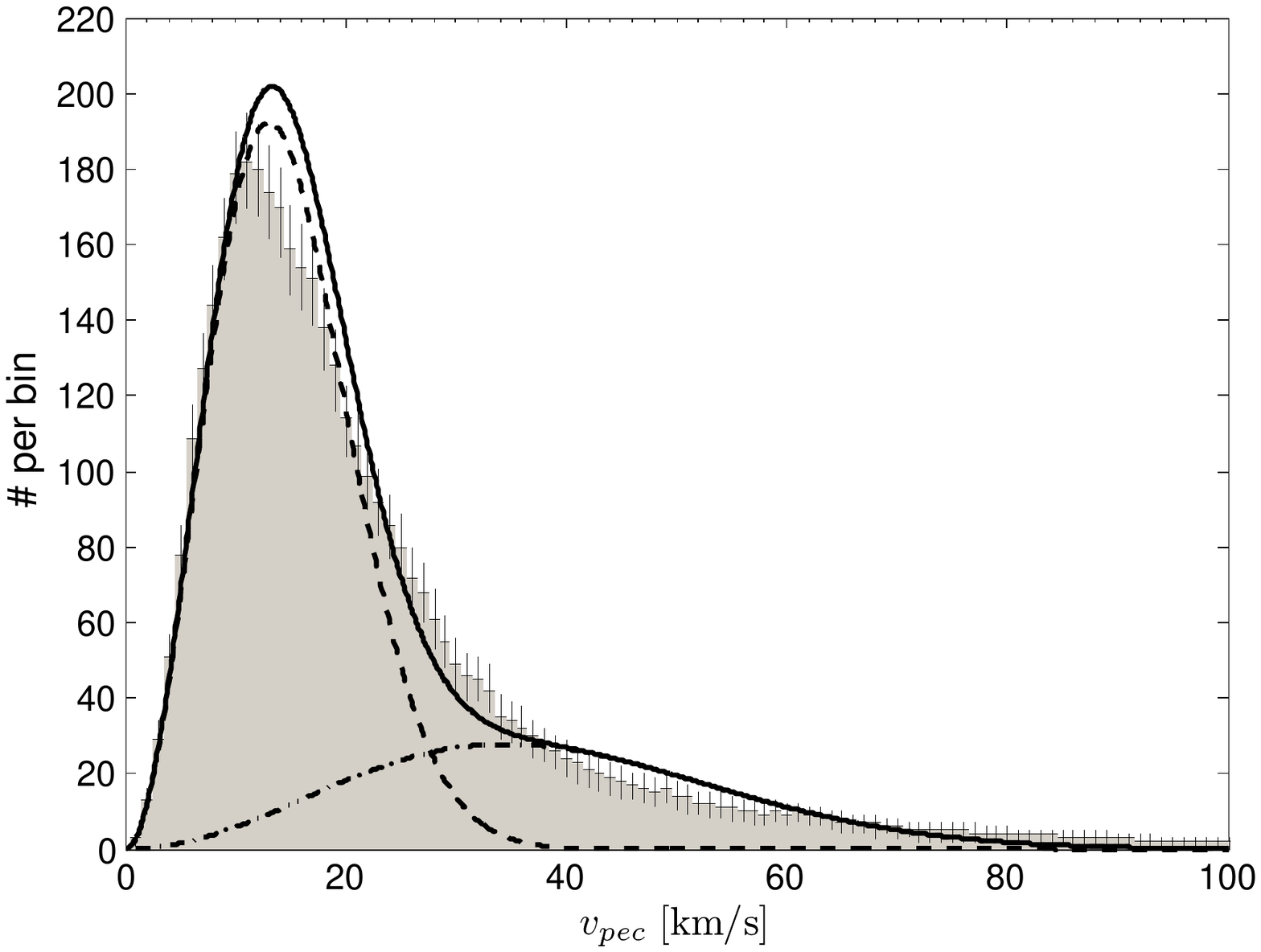}
\caption{Distribution of the peculiar 3D space velocity $v_{pec}$ (shaded histogram). The dashed curve shows the distribution for the low velocity group whereas the dashed-dotted curve is for the high velocity group. The two curves intersect at $v_{pec}=\unit[28]{km/s}$. The total distribution as the sum of the two is represented by the full line. Considering that different moving groups and associations of stars are present, thus both, the low and high velocity groups of stars, might be actually better represented by a superposition of many Maxwellians with slightly different parameters, the fit ($\chi^2_{red}=1{.}57$) is satisfying.}
\label{fig:2Maxwell}
\end{figure}

Runaway stars were firstly described as the stars that are responsible for the longer tail in the velocity distribution such that it is not sufficiently describable with one Maxwellian distribution \citep{1961BAN....15..265B}. \citet{1979ApJ...232..520S} generalised the definition such that runaway stars are the members of the so-called high velocity group. These are stars with large peculiar velocities that can be represented by an additional Maxwellian distribution. The other Maxwellian distribution incorporates stars with lower velocities, thus the low velocity group (normal Population I stars). He pointed out that by applying a velocity cutoff to identify the runaway stars, a certain fraction of them would be missed. \\
However, this issue can only be handled for the determination of space frequencies \citep[see][]{1991AJ....102..333S} and a velocity cutoff is still inevitable for the identification of runaway star candidates. To obtain a reasonable cutoff, we fit the distribution of the peculiar space velocities $v_{pec}$ of the sample stars (4180 with full kinematics) with two Maxwellians (\autoref{fig:2Maxwell}). We evaluate the velocity errors utilising a Monte-Carlo simulation varying $\pi$, $\mu_\alpha^\ast$, $\mu_\delta$ and $v_r$ within their confidence intervals.\footnote{Note that the velocity errors are not symmetric due to the inverse proportionality regarding $\pi$.} We find that
\begin{eqnarray}
& \sigma_L  & = \unit[9{.}2\pm0{.}2]{km/s},\notag\\
& \sigma_H & = \unit[24{.}4\pm1{.}2]{km/s},\\
& f_H & = \unit[27{.}7\pm1{.}9]{\%},\notag
\end{eqnarray}
where $\sigma_L$ and $\sigma_H$ are the average velocity dispersions of the low and high velocity groups, respectively, and $f_H$ is the relative frequency of the high velocity group. The derived dispersions are in agreement with those found by \citet{1979ApJ...232..520S} (with $\sigma_H$ being slightly smaller) whereas $f_H$ is smaller; however published values for $f_H$ vary from $\unit[34\pm14]{\%}$ (\citealt{1961BAN....15..265B}, corrected -- see \citealt{1991AJ....102..333S}) to $\unit[55\pm12]{\%}$ \citep{1979ApJ...232..520S}. Furthermore, Stone's star sample contains a much smaller number of stars than ours. In addition, to make sure that the low mass stars in our sample do not distort the results, we check whether the outcome differs from a subsample comprising only O and B type stars as well as Wolf-Rayet stars (2368 stars with full kinematics). Since we do not find significant differences we conclude that young stars, no matter if of low or high mass, share the same kinematic properties.
The Maxwellian functions of both groups intersect at $\unit[28]{km/s}$. Following \citet{1979ApJ...232..520S}, a star is a probable member of the high velocity group if 
\begin{equation}
v_{pec}>\unit[28]{km/s}.\label{eq:vgescrit}
\end{equation}
For that reason this will be our velocity cutoff defining runaway stars. In theory, with this definition we are able to identify 73 per cent of the high velocity group members while the contamination of low velocity stars is nine per cent.\\
We perform a Monte-Carlo simulation varying the observables $\pi$, $\mu_\alpha^\ast$, $\mu_\delta$ and $v_r$ within their uncertainty intervals and evaluate the probability of a star being a runaway star (the probabilities are given in \autoref{tab:runprob}). For 972 stars, the probability is higher than 50 per cent. Allowing for a nine per cent contamination of low velocity stars, this means a probability criterion of 50 per cent allows us to identify 78 per cent of the high velocity members.

%---------------------------------------------------------------

\subsection{Runaway stars identified from $U$, $V$, $W$, their radial and tangential velocities or proper motions}\label{subsec:other}

In addition to the peculiar 3D space velocities $v_{pec}$ we investigate their 1D components $U$, $V$ and $W$ separately to identify potentially slower high velocity group members which may show an exceptionally high velocity in only one direction. For the same reason we also investigate the peculiar radial velocities $v_{r,pec}$.\\
45 per cent of the stars in our sample do not have radial velocity measurements available. Among these cases, the only way to identify runaway candidates is to use their peculiar 2D tangential Galactic velocities $v_{t,pec}$ or its 1D components which are the peculiar proper motion in Galactic longitude $\mu_{l,pec}$ and Galactic latitude $\mu_{b,pec}$. To make the velocities comparable, we transfer the proper motions into 1D velocities $v_{l,pec} = 4.74\cdot\mu_{l,pec}/\pi$ and $v_{b,pec}=4.74\cdot\mu_{b,pec}/\pi$.\\

All velocity distributions contain the two velocity groups of stars (\autoref{subsec:3Dvel}) and can be fitted with bimodal functions (Gauss\-ians for the 1D cases $U$, $V$, $W$, $v_{r,pec}$, $v_{l,pec}$ and $v_{b,pec}$, 2D Maxwellians for the 2D case $v_{t,pec}$). \Autoref{tab:UVWRVvtvlvb} lists the fitting results adopting $f_H=\unit[27{.}7\pm1{.}9]{\%}$ (see \autoref{subsec:3Dvel}).\\
\begin{table*}
\centering
\caption{Fit results and curve intersection points for the different velocity components. The first Column gives the velocity component examined. Columns 2 and 4 give the centre velocities $v_{cL}$ and $v_{cH}$ (for the 1D cases, i.e. Gaussian fit) and Columns 3 and 5 give the velocity dispersions $\sigma_L$ and $\sigma_H$ of the low and high velocity groups, respectively. All errors are formal $1\sigma$ errors. In Column 6 the intersection points of the curves representing the low and high velocity groups are given (for the 1D cases two intersection points exist (negative and positive sides of the distribution), they are approximate since the distribution is not exactly symmetric). The last Column names the Figure in which the particular distribution is shown.}\label{tab:UVWRVvtvlvb}
\begin{tabular}{c d{1}<{\,\pm} >{\hspace{-0.1\tabcolsep}}d{3} d{1}<{\,\pm} >{\hspace{-0.1\tabcolsep}}d{3} d{1}<{\,\pm} >{\hspace{-0.1\tabcolsep}}d{3} d{1}<{\,\pm} >{\hspace{-0.1\tabcolsep}}d{3} c c}
\toprule
& \multicolumn{2}{c}{$v_{cL}$ [km/s]} & \multicolumn{2}{c}{$\sigma_L$ [km/s]}	&	\multicolumn{2}{c}{$v_{cH}$ [km/s]} & \multicolumn{2}{c}{$\sigma_H$ [km/s]} &  \multicolumn{1}{c}{intersection [km/s]} & Fig.\\\midrule
$U$ & -0,3 & 0,3 & 10,7 & 0,3 & 4,1 & 1,5 & 26,3 & 1,0  & $\pm23$ & \ref{subfig:U2Gauss}\\ 
$V$ & 0,0 & 0,2 & 10,7 & 0,1 & -4,7 & 1,0 & 24,2 & 1,0  & $\pm23$ & \ref{subfig:V2Gauss} \\
$W$ & 0,0 & 0,1 & 5,3 & 0,1 & -3,8 & 0,5 & 17,3 & 0,6   & $\pm12$ & \ref{subfig:W2Gauss} \\
$v_{r,pec}$ & 0,2 & 0,3 &  11,9 & 0,2 & -4,9 & 1,1 & 28,5 & 0,9 & $\pm25$ & \ref{fig:RV2Gauss} \\
$v_{t,pec}$ & \multicolumn{2}{c}{--} & 7,4 & 0,1 & \multicolumn{2}{c}{--} & 21,7 & 0,6  & $20$ & \ref{fig:2Maxwell_vtrans}\\
$v_{l,pec}$ & -2,7 & 0,1 & 8,9 & 0,1 & -2,6 & 0,7 & 27,3 & 0,8 & $\pm19$ & \ref{subfig:l2Gauss}\\
$v_{b,pec}$ & 0,6 & 0,1 & 5,3 & 0,1 & -3,3 & 0,4 & 18,3 & 0,7 & $\pm11$ & \ref{subfig:b2Gauss}\\
\bottomrule
\end{tabular}
\end{table*}
\begin{figure*}
\centering
\subfigure[$U$, $\chi^2_{red}=0{.}73$]{\includegraphics[width=0.33\textwidth, viewport= 40 200 535 610]{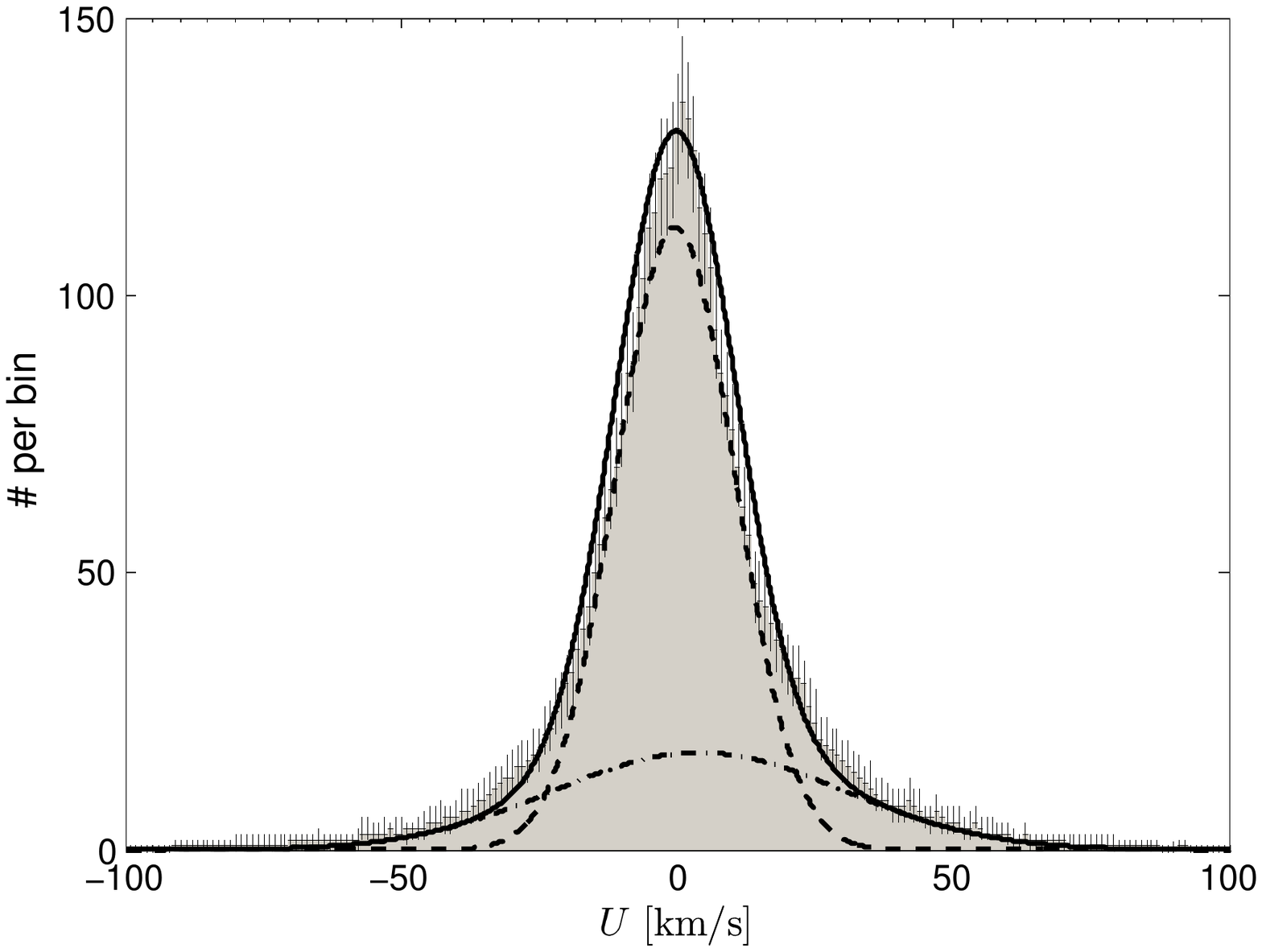}\label{subfig:U2Gauss}}\nolinebreak
\subfigure[$V$, $\chi^2_{red}=0{.}55$]{\includegraphics[width=0.33\textwidth, viewport= 40 200 535 610]{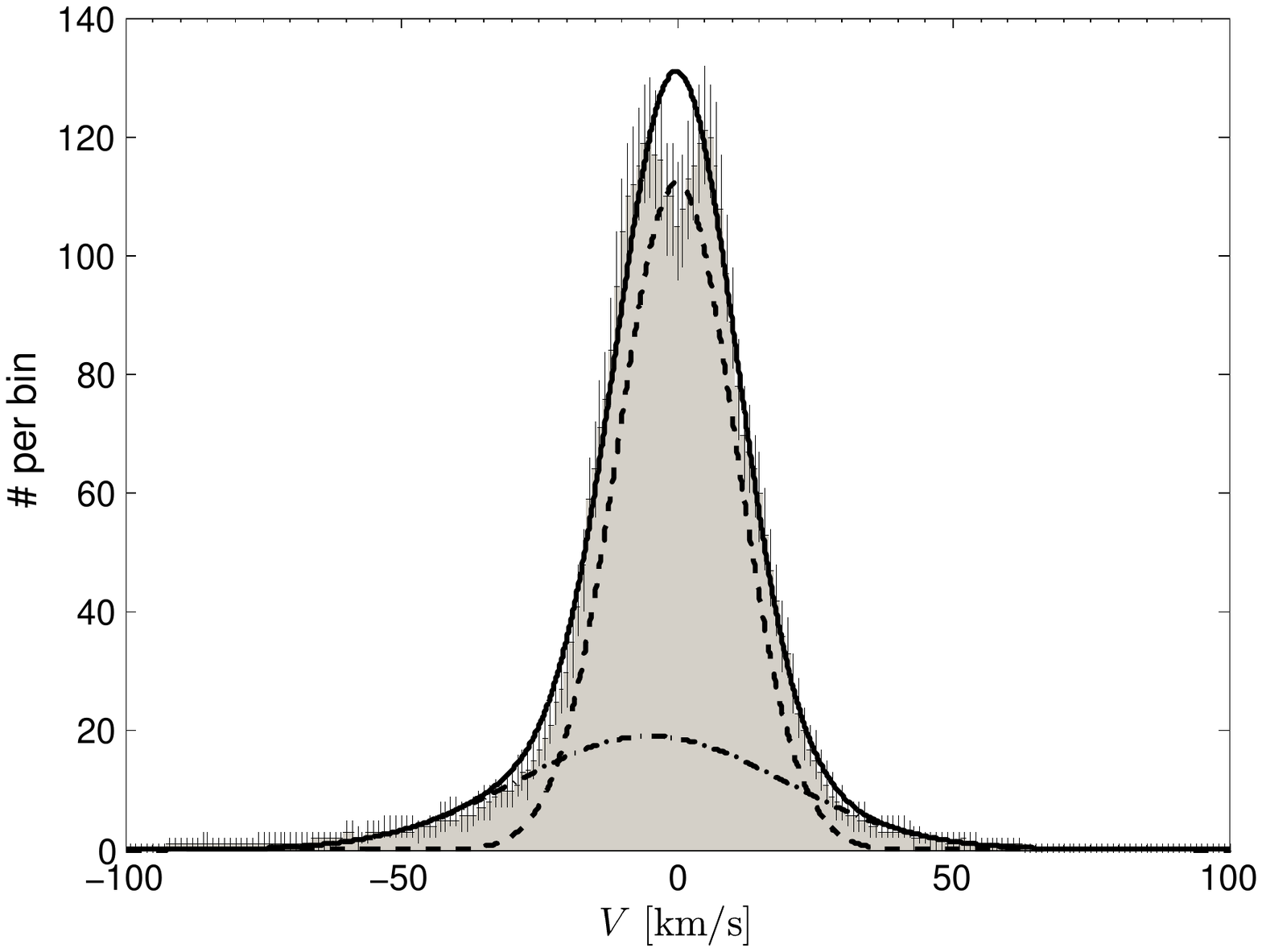}\label{subfig:V2Gauss}}\nolinebreak
\subfigure[$W$, $\chi^2_{red}=0{.}53$]{\includegraphics[width=0.33\textwidth, viewport= 40 200 535 610]{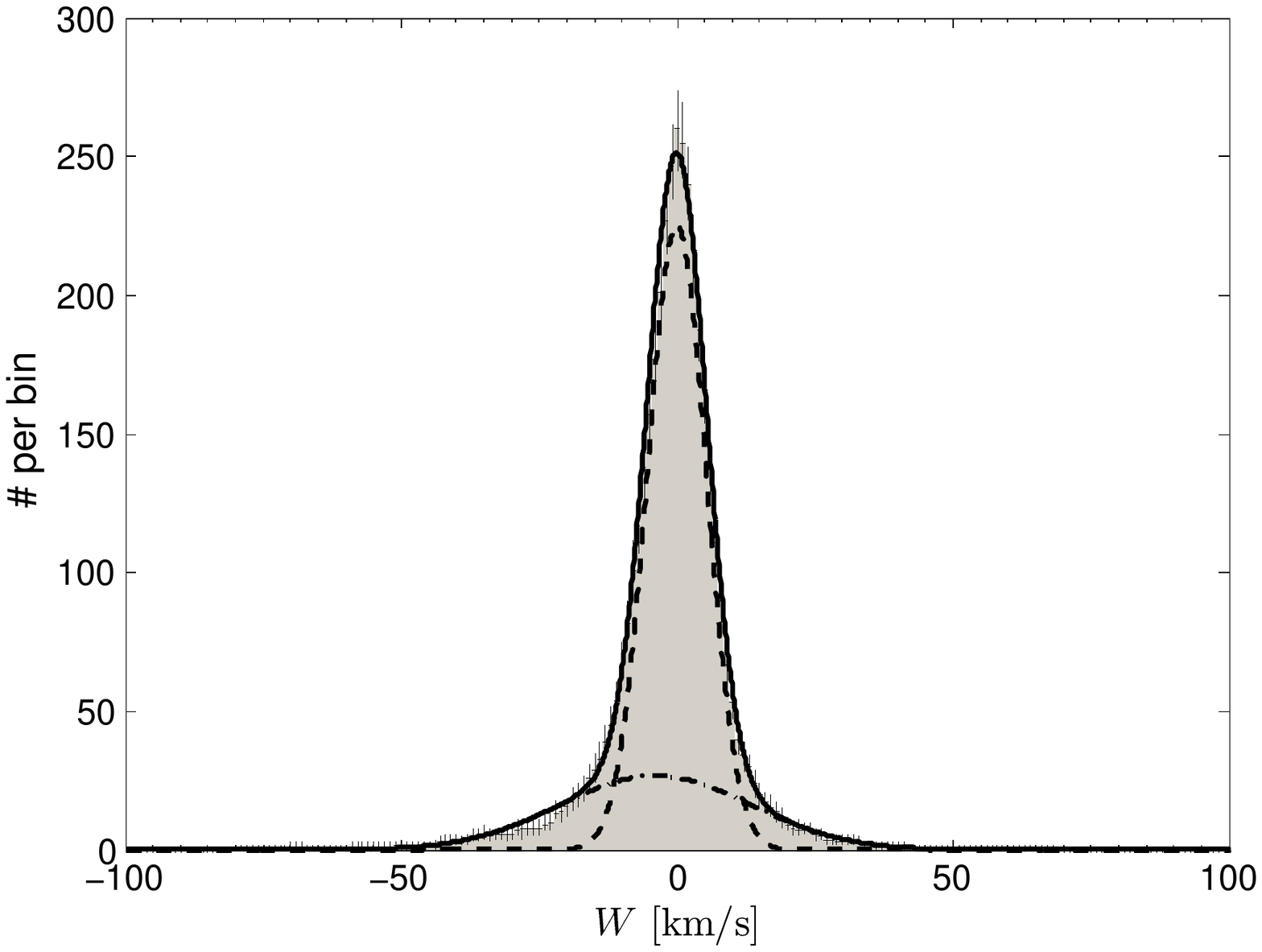}\label{subfig:W2Gauss}}
\caption{Distribution of $U$, $V$ and $W$ (shaded histograms) -- the 1D components of $v_{pec}$. The dashed curves represent the low velocity group while the dashed-dotted curves display the high velocity group. The two curves intersect at $U\approx\unit[\pm23]{km/s}$, $V\approx\unit[\pm23]{km/s}$ and $W\approx\unit[\pm12]{km/s}$, respectively. The total distribution as the sum of the two is represented by the full line.}
\label{figs:UVW2Gauss}
\end{figure*}
\begin{figure}
\centering
\includegraphics[width=0.45\textwidth, viewport= 40 200 535 610]{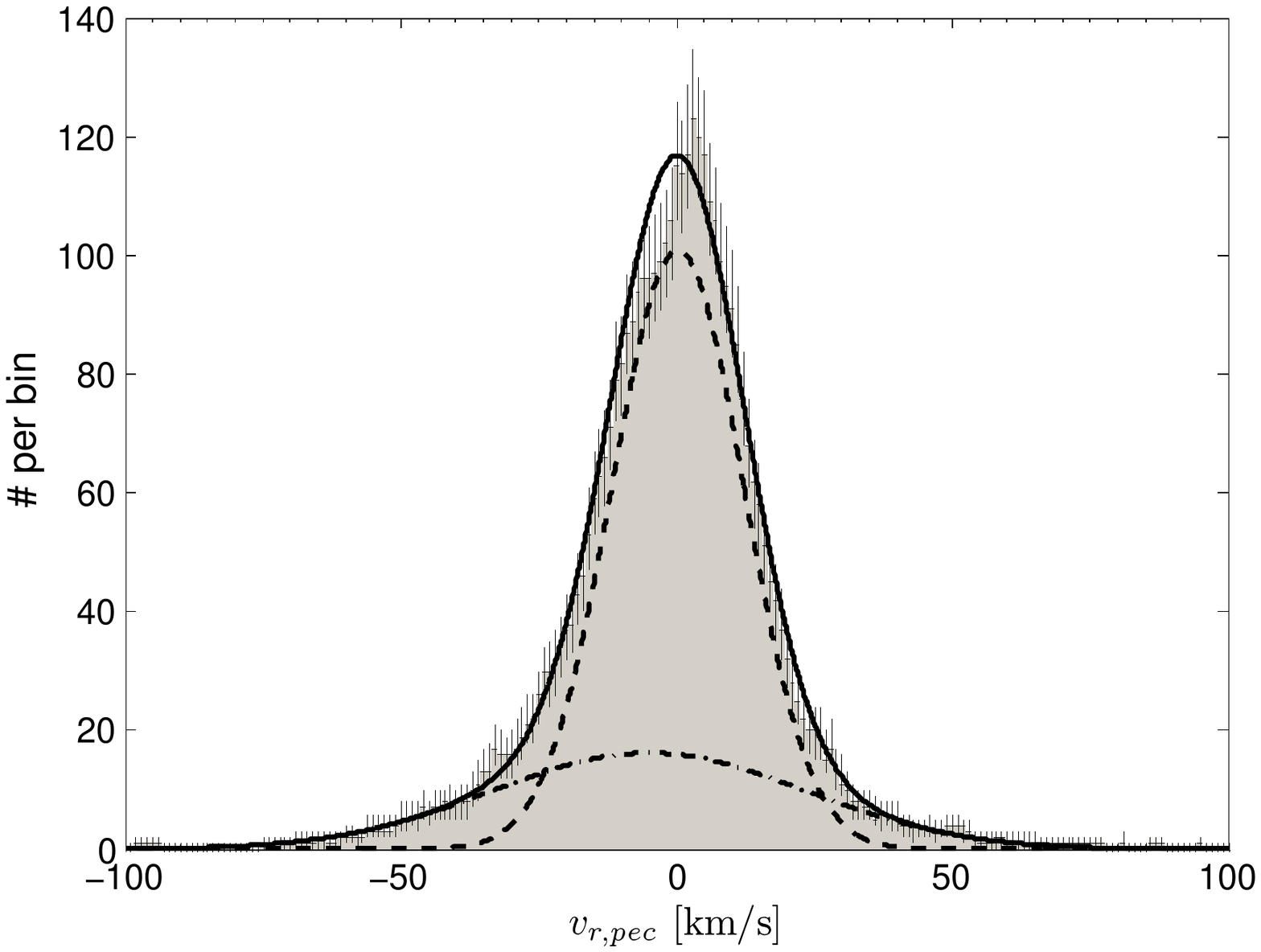}
\caption{Distribution of the peculiar 1D radial velocity $v_{r,pec}$ (shaded histogram). The dashed curve represents the low velocity group while the dashed-dotted curve displays the high velocity group. The two curves intersect at $v_{r,pec}\approx\unit[\pm25]{km/s}$. The total distribution as the sum of the two is represented by the full line ($\chi^2_{red}=0{.}33$).}
\label{fig:RV2Gauss}
\end{figure}
\begin{figure}
\centering
\includegraphics[width=0.45\textwidth, viewport= 40 200 535 610]{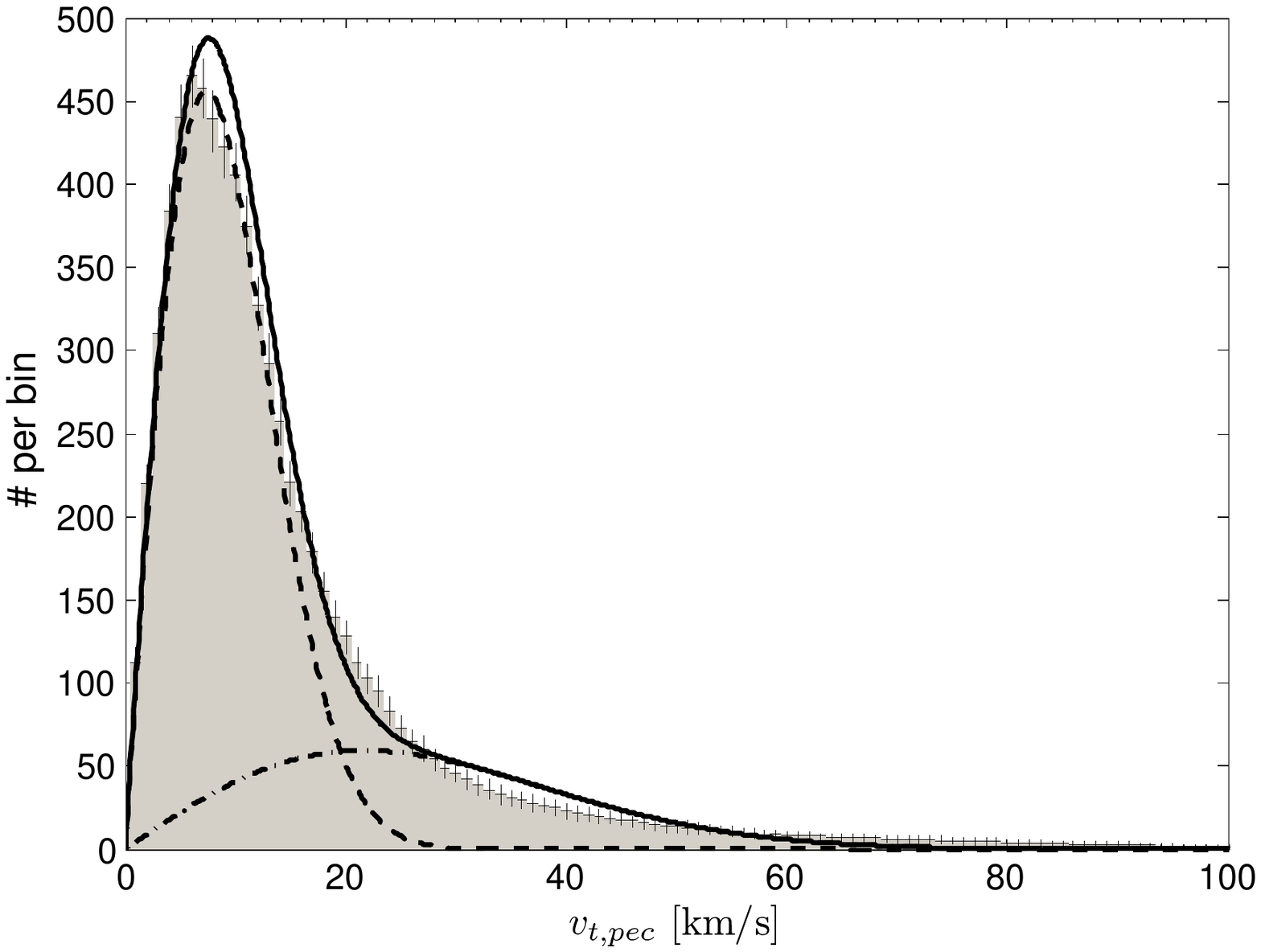}
\caption{Distribution of the peculiar 2D tangential Galactic velocity $v_{t,pec}$ (shaded histogram). The dashed curve represents the low velocity group while the dashed-dotted curve displays the high velocity group. The two curves intersect at $v_{r,pec}=\unit[20]{km/s}$. The total distribution as the sum of the two is represented by the full line ($\chi^2_{red}=2{.}58$, see comment to \autoref{fig:2Maxwell}).}
\label{fig:2Maxwell_vtrans}
\end{figure}
\begin{figure}
\centering
\subfigure[$v_{l,pec}$, $\chi^2_{red}=0{.}47$]{\includegraphics[width=0.25\textwidth, viewport= 40 200 535 610]{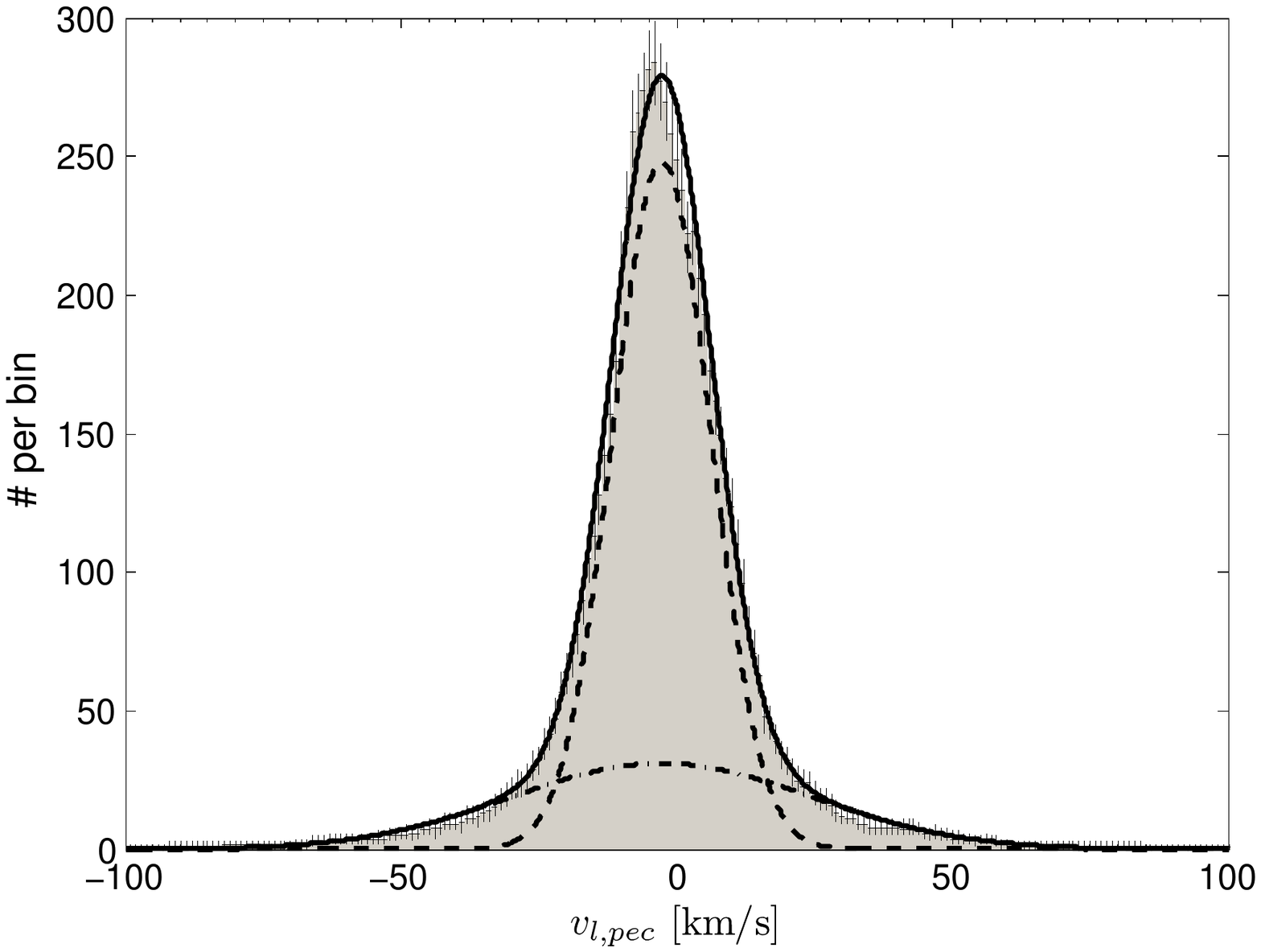}\label{subfig:l2Gauss}}\nolinebreak
\subfigure[$v_{b,pec}$, $\chi^2_{red}=1{.}06$]{\includegraphics[width=0.25\textwidth, viewport= 40 200 535 610]{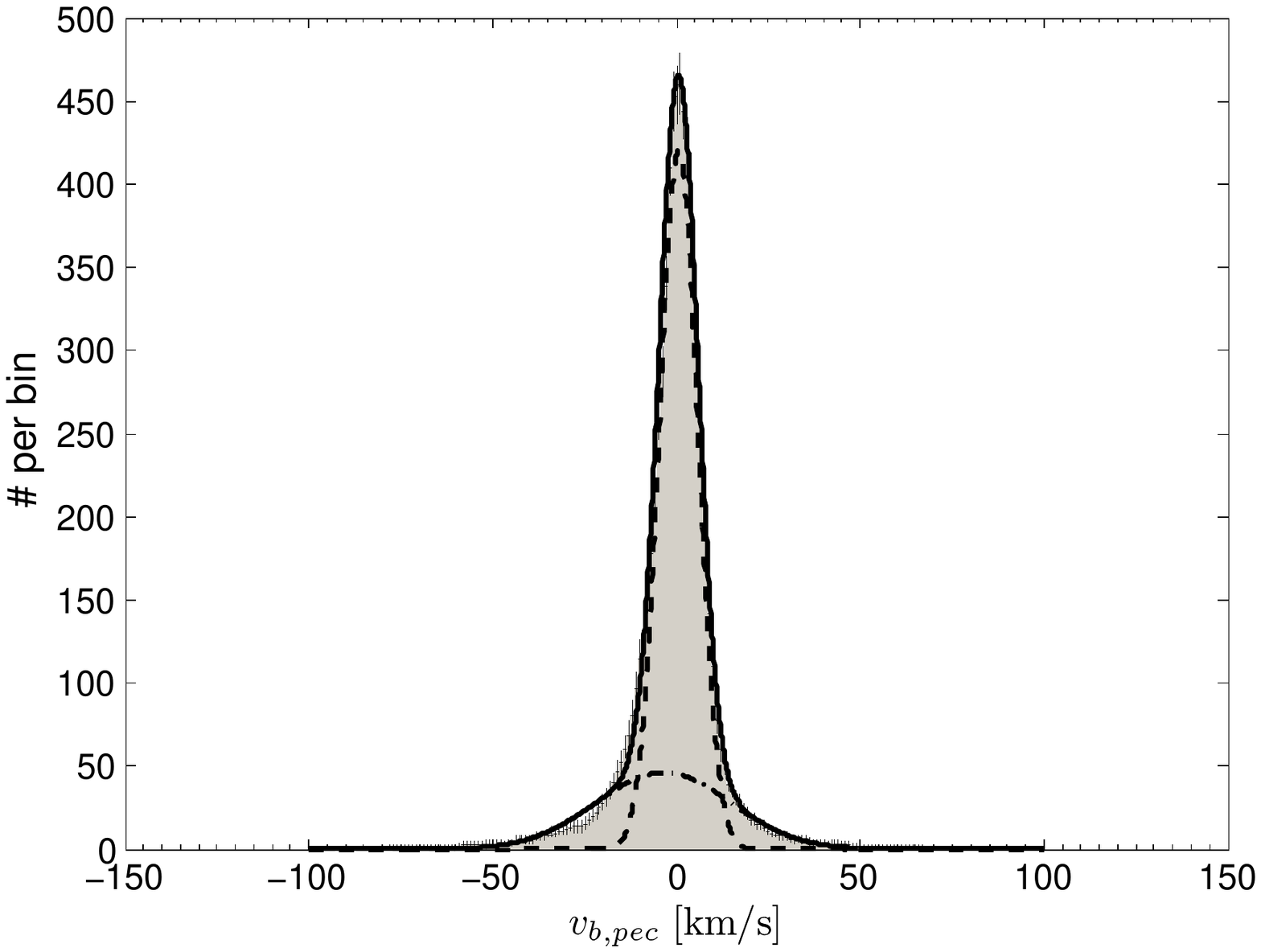}\label{subfig:b2Gauss}}
\caption{Distribution of the 1D components of the peculiar 2D tangential Galactic velocity $v_{t,pec}$: $v_{l,pec}$ and $v_{b,pec}$ (shaded histograms). The dashed curves represent the low velocity group while the dashed-dotted curves display the high velocity group. The two curves intersect at $v_{lpec}\approx\unit[\pm19]{km/s}$ and $v_{b,pec}\approx\unit[\pm11]{km/s}$, respectively. The total distribution as the sum of the two is represented by the full line.}
\label{figs:2Gauss_pmpec}
\end{figure}
The velocity dispersions of the high velocity group are consistent with an isotropic velocity distribution arising from the runaway producing mechanisms (see \autoref{appsec:adddisp}). Moreover, the low velocity group dispersions are in good agreement with those of young disk stars \citep[e.g.][]{1965gast.conf...61D,1981gask.book.....M}.\\
Since the velocity distribution of the low velocity group is not isotropic the one of the high velocity group cannot be isotropic either (see \autoref{appsec:adddisp}). Thus, we cannot simply translate the criterion given by \autoref{eq:vgescrit} into the 1D case ($\left|X\right|>\unit[28/\sqrt{3}]{km/s}$, where $X=U, V, W, v_{r,pec}, v_{l,pec}$ or $v_{b,pec}$). In addition, such a 1D criterion would lead to a contamination of low velocity group stars among the identifications of approximately 50 per cent e.g. in the $U$ and $V$ components.\\
For those reasons, as for the peculiar spatial velocity $v_{pec}$, we choose the intersection points of the curves also for the 1D cases as well as $v_{t,pec}$ to define the runaway criteria. In \autoref{tab:intexp}, we specify the selection criteria, theoretical expectations concerning the possible identifications as well as the number of runaway star candidates identified (stars with a runaway probability higher than 50 per cent).
\begin{table}
\centering
\caption{Runaway selection criteria $|v|>v_{crit}$ for the different velocity components. The limits correspond to the intersection points of the curves representing the low and high velocity groups. Columns 3 and 4 give the theoretical expectations concerning possible identifications of high velocity group members (the fraction of high velocity group stars $f_{id,th}$ satisfying the selection criteria) and the contamination of low velocity group members (the fraction of low velocity group stars $f_{c,th}$ that also satisfy the selection criteria) while in the last two Columns the number of runaway star candidates $N$ and the number of new identifications $N_{new}$ (compared to \autoref{subsec:3Dvel} and previous lines in the table) is quoted. Note that for $U$, $V$, $W$ and $v_{r,pec}$ only 4180 stars could be analysed whereas for $v_{t,pec}$, $v_{l,pec}$ and $v_{b,pec}$ the whole sample of 7663 stars was used.}\label{tab:intexp}
\begin{tabular}{c c c c c c}
\toprule
   &  \multicolumn{1}{c}{$v_{crit}$ [km/s]} & $f_{id,th}$ [\%] & $f_{c,th}$ [\%] & $N$ & $N_{new}$\\\midrule
$\left|U\right|$         & $23$ & $39$ & $17$ & $776$  & $70$\\ 
$\left|V\right|$         & $23$ & $35$ & $19$ & $452$  & $43$\\
$\left|W\right|$         & $12$ & $50$ & $12$ & $609$  & $190$\\
$\left|v_{r,pec}\right|$ & $25$ & $39$ & $20$ & $588$  & $12$\\
$v_{t,pec}$              & $20$ & $66$ & $9$ & $1513$ & $768^\mathrm{a}$\\
$\left|v_{l,pec}\right|$ & $19$ & $49$ & $18$ & $1162$ & $33^\mathrm{b}$\\
$\left|v_{b,pec}\right|$ & $11$ & $56$ & $15$ & $1266$ & $265^\mathrm{c}$\\
\bottomrule
\multicolumn{6}{l}{$^\mathrm{a}$ 643 of them without $v_r$ measurements}\\
\multicolumn{6}{l}{$^\mathrm{b}$ 33 of them without $v_r$ measurements}\\
\multicolumn{6}{l}{$^\mathrm{c}$ 265 of them without $v_r$ measurements}
\end{tabular}
\end{table}
Allowing for the specific contamination of low velocity group members, the number of identifications is generally in agreement with our theoretical predictions (\autoref{tab:intexp}).\\

With 972 runaway star candidates found in \autoref{subsec:3Dvel} and 1381 identifications in \autoref{tab:intexp} ($N_{new}$), we find a total of 2353 runaway star candidates with a runaway probability higher than 50 per cent regarding at least one velocity investigated (\autoref{tab:runprob}, the full table will be available via VizieR soon after publication).
\begin{table*}
\centering
\caption{Runaway probabilities for 2353 runaway star candidates as found in the previous section. Columns 3 to 10 list the individual probabilities $P$ for each velocity component. We regard stars with $P\geq0{.}50$ in at least one velocity as runaways. The peculiar space velocities $v_{pec}$ and peculiar tangential velocities $v_{t,pec}$ are given in the last two Columns. Here the first five entries are shown, the full table will be available via VizieR soon after publication.}\label{tab:runprob}
\begin{tabular}{r c d{2.2} d{2.2} d{2.2} d{2.2} d{2.2} d{2.2} d{2.2} d{2.2} >{$}r<{$} >{$}r<{$}}
\toprule
\multicolumn{1}{c}{HIP} & \multicolumn{1}{c}{other name} & \multicolumn{1}{c}{$P_{v_{pec}}$} & \multicolumn{1}{c}{$P_{U}$} & \multicolumn{1}{c}{$P_V$} & \multicolumn{1}{c}{$P_W$} & \multicolumn{1}{c}{$P_{v_{r,pec}}$} & \multicolumn{1}{c}{$P_{v_{t,pec}}$} & \multicolumn{1}{c}{$P_{v_{l,pec}}$} & \multicolumn{1}{c}{$P_{v_{b,pec}}$} & \multicolumn{1}{c}{$v_{pec}$ [km/s]} & \multicolumn{1}{c}{$v_{t,pec}$ [km/s]}\\\midrule
85 & CD-25 16747 & 0,89 & 0,93 & 0,47 & 0,22 & 0,01 & 0,96 & 0,52 & 0,99 & 52{.}9^{+22{.}8}_{-35{.}2} & 52{.}7^{+24{.}8}_{-37{.}2}\\
135 & HD 224908 & \multicolumn{1}{c}{--} & \multicolumn{1}{c}{--} & \multicolumn{1}{c}{--} & \multicolumn{1}{c}{--} & \multicolumn{1}{c}{--} & 0,99 & 0,00 & 1,00 & \multicolumn{1}{c}{--} & 22{.}1^{+1{.}9}_{-2{.}1}\\
145 & 29 Psc & 0,13 & 0,00 & 0,00 & 0,95 & 0,31 & 0,00 & 0,00 & 0,00 & 22{.}4^{+3{.}9}_{-4{.}1} & 1{.}9^{+2{.}0}_{-2{.}0}\\
174 & HD 240475 & 1,00 & 0,52 & 1,00 & 0,00 & 1,00 & 0,04 & 0,04 & 0,00 & 59{.}6^{+4{.}8}_{-5{.}2} & 6{.}9^{+3{.}0}_{-1{.}0}\\
278 & HD 225095 & 0,45 & 0,13 & 0,23 & 0,04 & 0,51 & 0,03 & 0,02 & 0,03 & 27{.}3^{+8{.}9}_{-7{.}1} & 9{.}5^{+3{.}0}_{-1{.}0}\\\bottomrule
\end{tabular}
\end{table*}

%---------------------------------------------------------------

\subsection{Stars with higher peculiar velocities compared to their neighbourhood or surrounding OB association/ cluster}\label{subsec:assocident}

\begin{figure}
\centering
\includegraphics[width=0.45\textwidth]{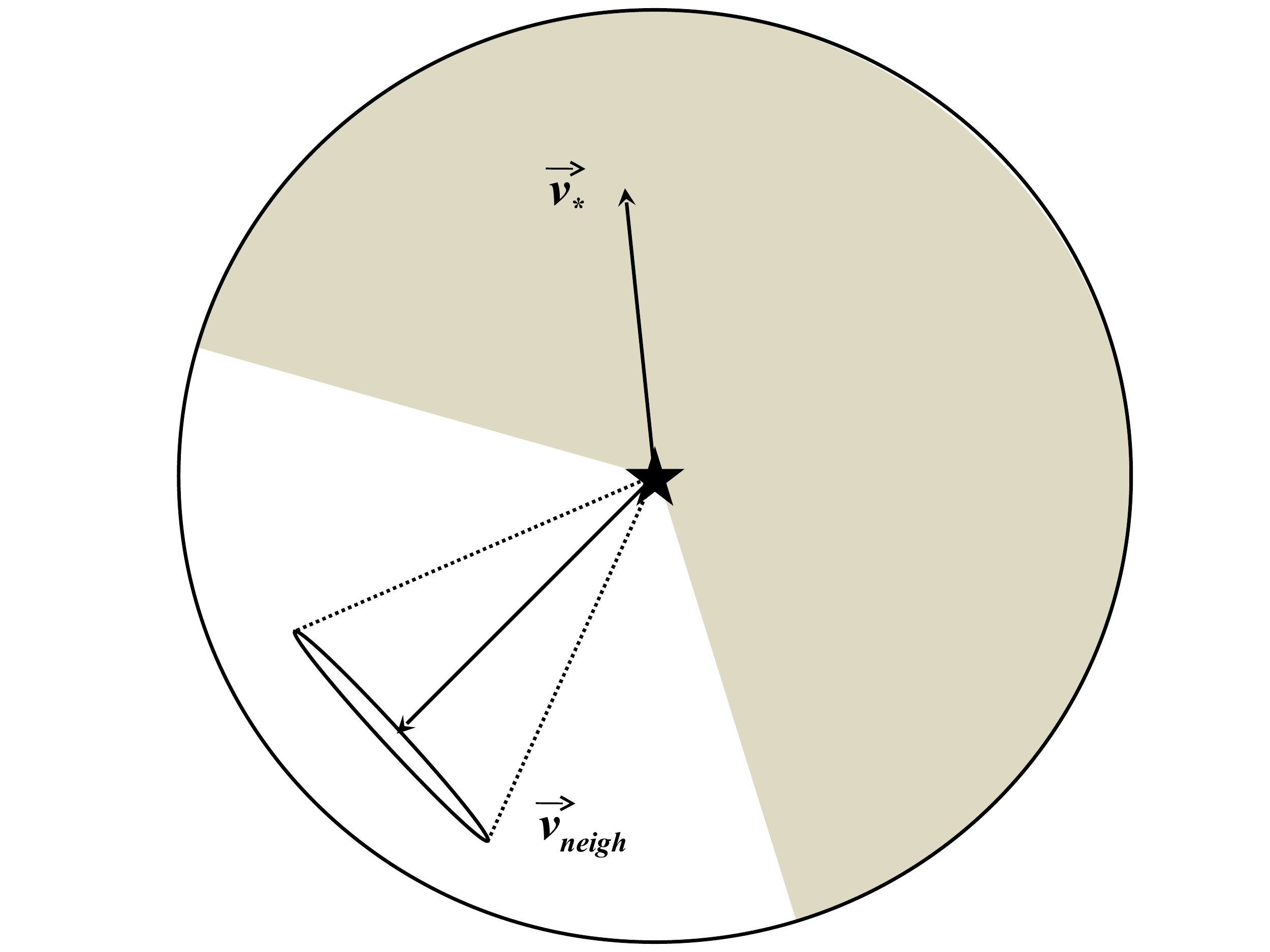}
\caption{Definition for identifying runaway stars by comparing with neighbouring stars. The neighbourhood is defined as a sphere around an individual star with a diameter of $24\,\mathrm{pc}$ (\autoref{footn:spherediam}) or as the OB association/ cluster (assumed to be spherical) inside which the star currently lies. The neighbourhood velocity $\vec{v}_{neigh}$ is given by the median velocity of the stars within the sphere or the mean motion of the association/ cluster member stars as listed in \citet{2010MNRAS.402.2369T}, respectively. If the velocity vector $\vec{v}_\star$ points into the grey shaded region, the star clearly moves away from its neighbouring stars and is hence a runaway star. The grey shaded area lies outside the $3\sigma$ error cone of $\vec{v}_{neigh}$ (dotted lines mark $1\sigma$). Note that the length of $\vec{v}_\star$ and $\vec{v}_{neigh}$ ($v_{pec}$ or $v_{t,pec}$) is not important here.}
\label{fig:Skizze_Kugel}
\end{figure}

Still, some high velocity group stars may not yet have been identified. For that reason we look for additional stars which show a different motion compared to their neighbouring stars. Since stars in clusters and associations share a common motion, runaway stars, i.e. stars that experienced some interaction (see \autoref{sec:intro}), can be identified through deviations from the common motion, especially if the velocity vector points towards a different direction than the cluster mean motion. From the previous investigations we can be sure that we identified all the runaway stars with high peculiar velocities. The most important criterion now is the direction of a star's velocity vector compared to its neighbouring stars.\\
We select a sphere with a diameter of $\unit[24]{pc}$\footnote{This is twice the median extension of all associations listed by \citet{2010MNRAS.402.2369T}.\label{footn:spherediam}} around each individual star to define its neighbourhood. All sample stars within this sphere are chosen as comparison stars. We calculate the vectors of the peculiar velocities $\vec{v}_{pec}=\left(U,V,W\right)$ and the peculiar tangential velocities $\vec{v}_{t,pec}=\left(v_{l,pec},v_{b,pec}\right)$ varying the observables within their confidence intervals. The neighbourhood velocity $\vec{v}_{neigh}$ is defined as the median velocity of the comparison stars. We define the runaway criterion such that $\vec{v}_\star$ must not lie within the $3\sigma$ error cone of $\vec{v}_{neigh}$ (\autoref{fig:Skizze_Kugel}). Varying the observables within their confidence intervals, we obtain runaway probabilities for each star (10000 Monte-Carlo runs).\\
\begin{figure}
\centering
\includegraphics[width=0.45\textwidth, viewport= 30 155 545 680]{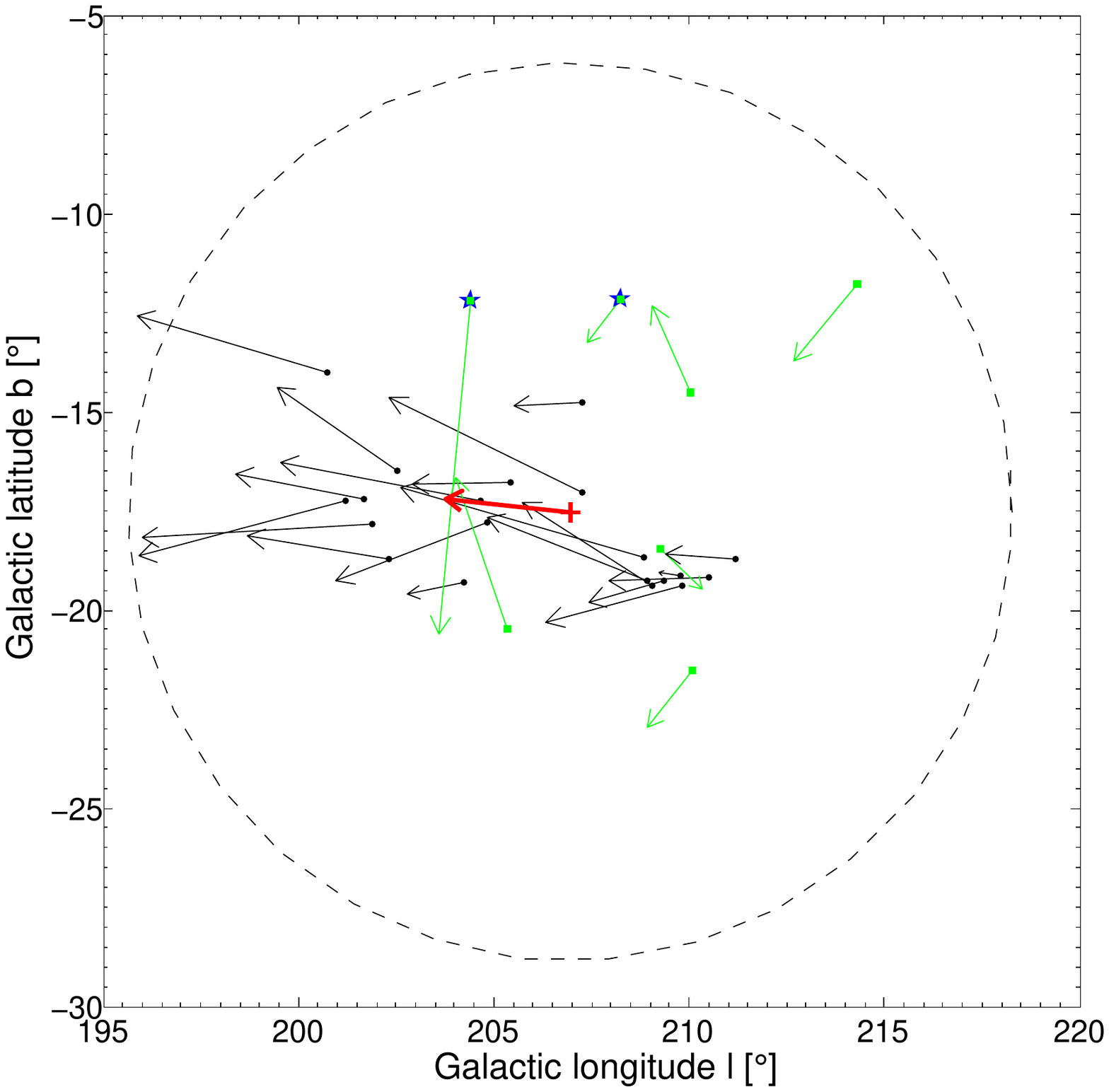}
\caption{Motion of Ori OB1 member stars. The red cross marks the centre of the association, the dotted ellipse its boundaries (assuming spherical shape). The red thick arrow shows the mean peculiar tangential motion of the whole association (\citealt{1989AJ.....98.1598B,1999osps.conf..411B,2001AstL...27...58D}; see also \citealt{2010MNRAS.402.2369T}), green squares are runaway star candidates satisfying the criterion defined by \autoref{fig:Skizze_Kugel} whereas blue stars mark runaway star candidates already defined by their large peculiar velocity (\autoref{sec:runaways}). The length of the arrows is scaled with distance to indicate tangential velocities.}
\label{fig:OriOB1}
\end{figure}
Examining $\vec{v}_{pec}$ and $\vec{v}_{t,pec}$, we find no stars with a probability of at least 50 per cent for being a runaway star under the above definition.\\

In addition, we compare the velocity vectors of each OB association and cluster listed in \citet{2010MNRAS.402.2369T}\footnote{We exclude Cas-Tau from the analysis owing to its large size \citep[cf.][their Table 2]{1999AJ....117..354D}.} with the velocity vectors of each individual star situated within the association boundaries as listed in \cite{2010MNRAS.402.2369T} (assuming the associations/ clusters are spherical, see \autoref{fig:OriOB1} for the Orion OB1 association as an example). From $\vec{v}_{pec}$, we identify 126 additional runaway star candidates. Another 58 runaway star candidates are identified from their $\vec{v}_{t,pec}$. All runaway star candidates (192 for $v_{pec}$, 221 for $v_{t,pec}$; 124 included in both) are listed in \autoref{tab:runcand3} (the full table will be available via VizieR soon after publication).
\begin{table*}
\caption{Runaway probabilities for runaway star candidates found by comparison with OB associations and clusters (as listed in \citealt{2010MNRAS.402.2369T}). 192 runaway stars are identified from $v_{pec}$ and 221 from $v_{t,pec}$. 124 stars are included in both lists. Columns 3 and 4 list the runaway probability $P$ as well as the association/ cluster. The absolute velocity values are given in columns 5 and 6. Errors correspond to 68 per cent confidence. The last Column indicates whether the star was already identified as a runaway star in \autorefs{subsec:3Dvel} and \ref{subsec:other} (``prev'' for previous identification; ``new'' for new identification). Here the first 5 entries for $\vec{v}_{pec}$ and $\vec{v}_{t,pec}$, respectively, are shown, the full table will be available via VizieR soon after publication.}\label{tab:runcand3}
\begin{tabular}{c c d{2.2} c >{$}r<{$} >{$}r<{$} l}
\toprule
\multicolumn{1}{c}{HIP} & \multicolumn{1}{c}{other name} & \multicolumn{1}{c}{$P$} & Assoc./ cluster & \multicolumn{1}{c}{$v_{pec}$ [km/s]} & \multicolumn{1}{c}{$v_{t,pec}$ [km/s]} & new ident.\\\midrule
\multicolumn{7}{c}{from $\vec{v}_{pec}$}\\\midrule
490	&	HD 105 &	1.00	&	$\beta$ Pic-Cap	&	11.0	^{+	2.9	}_{-	1.1	}&	10.6	^{+	2.9	}_{-	1.1	} & new\\
     	&	     	  &	0.99	&	ABDor	&	&	 & \\
1481	&	HD 1466 &	1.00	&	Tuc-Hor	&	10.8	^{+	2.9	}_{-	1.1	}&	10.7	^{+	2.9	}_{-	1.1	} & new\\
     	&	     	  &	1.00	&	$\beta$ Pic-Cap	&	&	& \\
     	&	     	  &	1.00	&	ABDor	&	&	& \\
1623	&	HD 1686  &	1.00	&	Tuc-Hor	&	27.7	^{+	1.9	}_{-	2.1	}&	27.5	^{+	2.9	}_{-	1.1	} & prev\\
     	&	     	  &	1.00	&	$\beta$ Pic-Cap	&	&	&\\
1803	&	BE Cet &	1.00	&	Tuc-Hor	&	27.4	^{+	2.9	}_{-	1.1	}&	27.1	^{+	2.9	}_{-	1.1	} & prev\\
     	&	     	  &	1.00	&	$\beta$ Pic-Cap	&	&	 & \\
     	&	     	  &	1.00	&	ABDor	&	&	 & \\
1910	&	GSC 08841-00065  &	0.65	&	$\beta$ Pic-Cap	&	13.1	^{+	2.9	}_{-	5.1	}&	13.0	^{+	2.9	}_{-	5.1	} & new\\\midrule
\multicolumn{7}{c}{from $\vec{v}_{t,pec}$}\\\midrule
135	&	HD 224908   &	0.96	&	ABDor	&	$--$&	22.1	^{+	1.9	}_{-	2.1	} & prev\\
439	&	HD 225213 &	1.00	&	Tuc-Hor	&	111.3	^{+	2.5	}_{-	1.5	}&	109.5	^{+	2.6	}_{-	1.4	} & prev\\
     	&	     	&	1.00	&	ABDor	&	&	& \\
490	&	HD 105   &	1.00	&	ABDor	&	11.0	^{+	2.9	}_{-	1.1	}&	10.6	^{+	2.9	}_{-	1.1	} & new\\
     	&	     	&	1.00	&	$\beta$ Pic-Cap	&	&	 & \\
544	&	V 439 And   &	1.00	&	Tuc-Hor	&	11.9	^{+	1.9	}_{-	2.1	}&	11.1	^{+	2.9	}_{-	1.1	} & new\\
     	&	     	&	1.00	&	$\beta$ Pic-Cap	&	&	 & \\
1128	&	HD 967&	1.00	&	Tuc-Hor	&	82.2	^{+	3.7	}_{-	4.3	}&	77.8	^{+	4.7	}_{-	5.3	} & prev\\
     	&	     	&	1.00	&	$\beta$ Pic-Cap	&	&	 &\\\bottomrule
\end{tabular}
\end{table*}
%

%------------------------------------------------------------------------

\subsection{Stars outside OB associations/ clusters and the Galactic plane}\label{subsec:outsideass}

As runaway stars were ejected from their birth site, i.e. its host OB association/ cluster or the Galactic plane, they are supposed to be isolated (outside any OB association/ cluster and the Galactic plane). Thus, we look for young stars that are clearly outside any OB association/ cluster listed in \citet{2010MNRAS.402.2369T} (outside three times the association radius which corresponds to approximately $3\sigma$) and probably outside the Galactic plane ($z>\unit[500]{pc}$\footnote{This number is derived from twice the low velocity dispersion in the $z$ direction ($\approx\unit[10]{km/s}$) and an age of $\unit[50]{Myr}$ (our age limit, see \autoref{subsec:agemass}). We do not use the individual ages of the stars owing to the large uncertainty inferred from different evolutionary models.\label{footn:Zcrit}}).\\
72 stars are situated well outside any OB association/ cluster and the Galactic plane; six of them were not identified as runaway star candidates in the previous sections. Those six are listed in \autoref{tab:Zcrit}. If they did not form in isolation they are runaway stars.
\begin{table*}
\centering
\caption{Additional young stars situated well outside any OB association/ cluster and the Galactic plane (see \autoref{subsec:outsideass} for criterion), i.e. runaway star candidates. Columns 3 to 4 give the distance $z$ to the Galactic plane as well as the star's velocity $W_{pec}$ in this direction. The absolute velocity values are given in columns 5 and 6. Errors correspond to 68 per cent confidence. Although some stars would be consistent with $z=0$ within $2\sigma$ and might be outliers we include them as potential runaway stars to not miss them only due to their large error bars.}\label{tab:Zcrit}
\begin{tabular}{c c >{$}r<{$} >{$}r<{$} >{$}r<{$} >{$}r<{$}}
\toprule
\multicolumn{1}{c}{HIP} & \multicolumn{1}{c}{other name} & \multicolumn{1}{c}{$z$ [pc]} & \multicolumn{1}{c}{$W_{pec}$ [km/s]} & \multicolumn{1}{c}{$v_{pec}$ [km/s]} & \multicolumn{1}{c}{$v_{t,pec}$ [km/s]}\\\midrule
5805 	& HD 7598 &	-477^{+280}_{-350}&	$--$ & $--$ &	13{.}1^{+12{.}9}_{-13{.}1} \\
11242 & HD 14920 &	-609^{+310}_{-255}	&	$--$ & $--$ &	15{.}9^{+9{.}9}_{-16{.}1} \\
50684	& RS Sex &	555^{+135}_{-245}	& -16{.}3^{+10{.}0}_{-8{.}0}	&	12{.}0^{+5{.}9}_{-10{.}1}&	6{.}6^{+5{.}0	}_{-7{.}0} \\
54769	& HD 97443 &	517^{+230}_{-265}	&	$--$ & $--$ &	5{.}7^{+4{.}0}_{-6{.}0} \\
56473	& 90 Leo &	488^{+320}_{-280}	& -3{.}2^{+6{.}0}_{-10{.}0}	& 14{.}3^{+7{.}9}_{-10{.}1}&	8{.}0	^{+6{.}0}_{-8{.}0}\\
70000 & HD 125504 & 544^{+300}_{-290}	& -10{.}2^{+8{.}0}_{-2{.}0}	& 25{.}4^{+6{.}9}_{-9{.}1}&	20{.}0	^{+3{.}9}_{-8{.}1} \\\bottomrule
\end{tabular}
\end{table*}
%
%------------------------------------------------------------------------

\subsection{Comparison with other authors}\label{subsec:otherauthors}

We compare our sample of runaway star candidates with lists from other authors \citep[most important sources:][]{1961BAN....15..265B,1974ApJ...190..653B,1974RMxAA...1..211C,1979ApJ...232..520S,1986ApJS...61..419G,1990AJ.....99..608L,1990AA...236..357C,1996AJ....111.1220P,1999AA...345..321M,2001AA...365...49H,2005AA...431L...1M,2005AA...437..247D,2006AJ....131.3047M}. 24 proposed runaway candidates which satisfy our initial sample criteria (HIP star, $\pi-\sigma_\pi\leq\unit[1/3]{mas}$, \autoref{eq:taulimit}), i.e. are contained in our young star sample, were not identified as runaway stars by us. They are listed in \autoref{tab:missingrun} along with the respective publication source.\\
\begin{table}
\centering
\caption{Runaway star candidates from Hipparcos proposed in the literature that are absent from our sample of 2543 runaway stars.}\label{tab:missingrun}
\begin{tabular}{c c}
\toprule
HIP & Ref.\\\midrule
102195 & \citet{1961BAN....15..265B,1974ApJ...190..653B}\\
35412 & \citet{1974RMxAA...1..211C}\\
40328 & \citet{1974RMxAA...1..211C}\\
81736 & \citet{1974RMxAA...1..211C}\\
18246 & \citet{1979ApJ...232..520S}\\
100214 & \citet{1979ApJ...232..520S}\\
114104 & \citet{1979ApJ...232..520S}\\
117221 & \citet{1979ApJ...232..520S}\\
67279 & \citet{1990AJ.....99..608L}\\
3881 & \citet{2001AA...365...49H}\\
20330 & \citet{2001AA...365...49H}\\
38455 & \citet{2001AA...365...49H}\\
48943 & \citet{2001AA...365...49H}\\
86768 & \citet{2001AA...365...49H}\\
92609 & \citet{2001AA...365...49H}\\
103206 & \citet{2001AA...365...49H}\\
69640 & \citet{2005AA...431L...1M}\\
73720 & \citet{2005AA...431L...1M}\\
80338 & \citet{2005AA...431L...1M}\\
90074 & \citet{2005AA...431L...1M}\\
104361 & \citet{2005AA...431L...1M}\\
1415 & \citet{2005AA...437..247D}\\
37074 & \citet{2005AA...437..247D}\\
97757 & \citet{2005AA...437..247D}\\
\bottomrule
\end{tabular}
\end{table}
We add three of the missing stars, namely HIP 3881 (=35 And), HIP 48943 (=OY Hya) and HIP 102195 (=V4568 Cyg) as well as HIP 26241 (=$\iota$ Ori) due to its DES origin (see individual discussion in \autoref{appsec:otherauthors}).\\
As indicated by the discussion in \autoref{appsec:otherauthors}, several problems may lead to mis- or non-identification of runaway stars. The major issue here is certainly the distance of a star that highly affects the calculated velocities. Since we used precise Hipparcos parallaxes (while previous studies often used ground-based distances) our results are not significantly influenced by that. Moreover, we derive runaway probabilities accounting for the errors on all observables instead of evaluating only a single velocity. \\
Another problem arises from the multiplicity of stars, e.g. 390 stars in our young star sample are spectroscopic binaries \citep{2009yCat....102020P,1988A&AS...75..441P}. We obtained radial velocities from catalogues which typically list the systemic radial velocity.

%______________________________________________________________

\section{Summary and conclusions}\label{sec:conclusions}

\begin{figure}
\centering
\subfigure[$v_{t,pec}$]{\includegraphics[width=0.25\textwidth, viewport= 40 200 535 610]{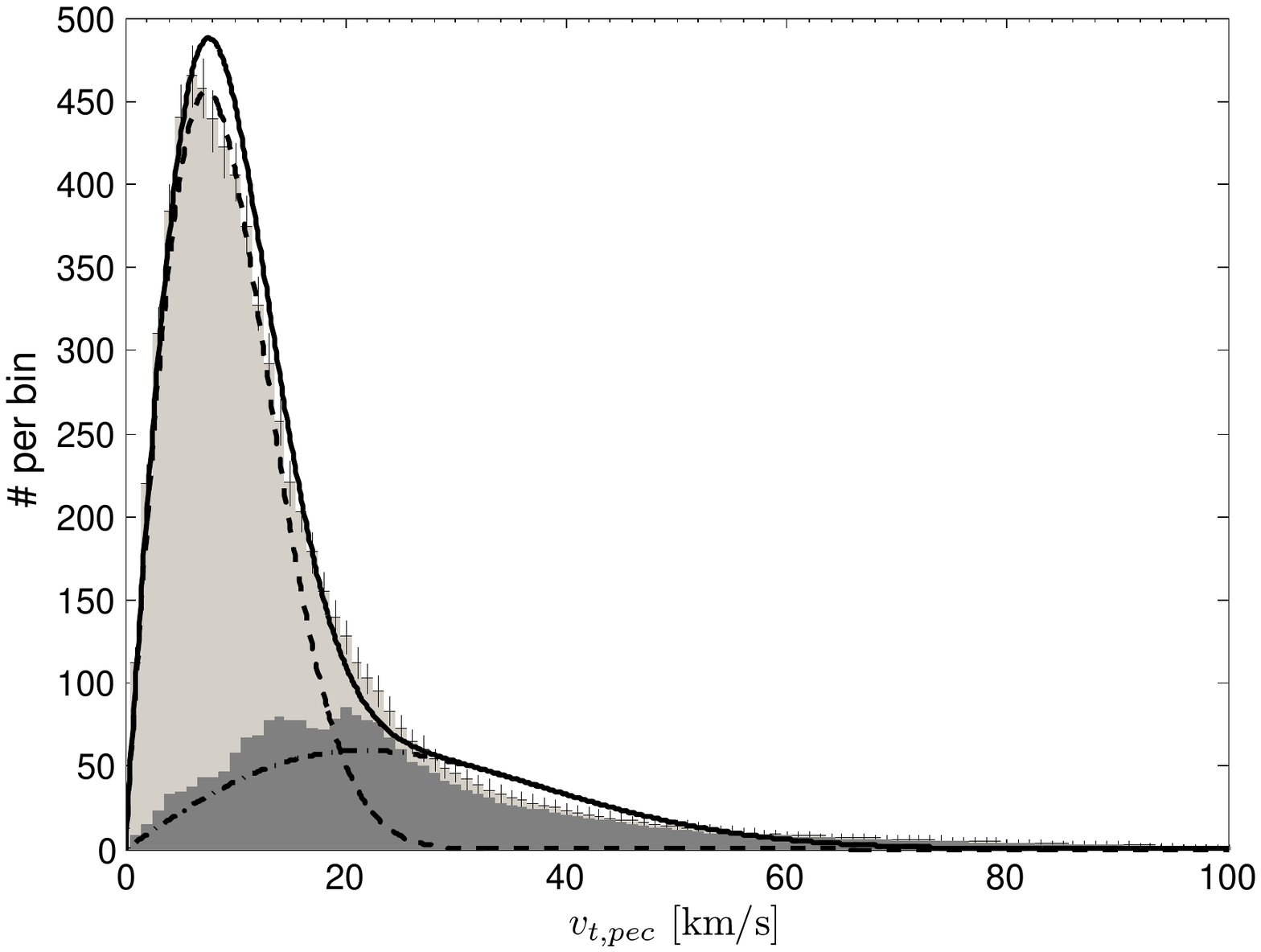}\label{subfig:vt_mitrun}}\nolinebreak
\subfigure[$v_{pec}$]{\includegraphics[width=0.25\textwidth, viewport= 40 200 535 610]{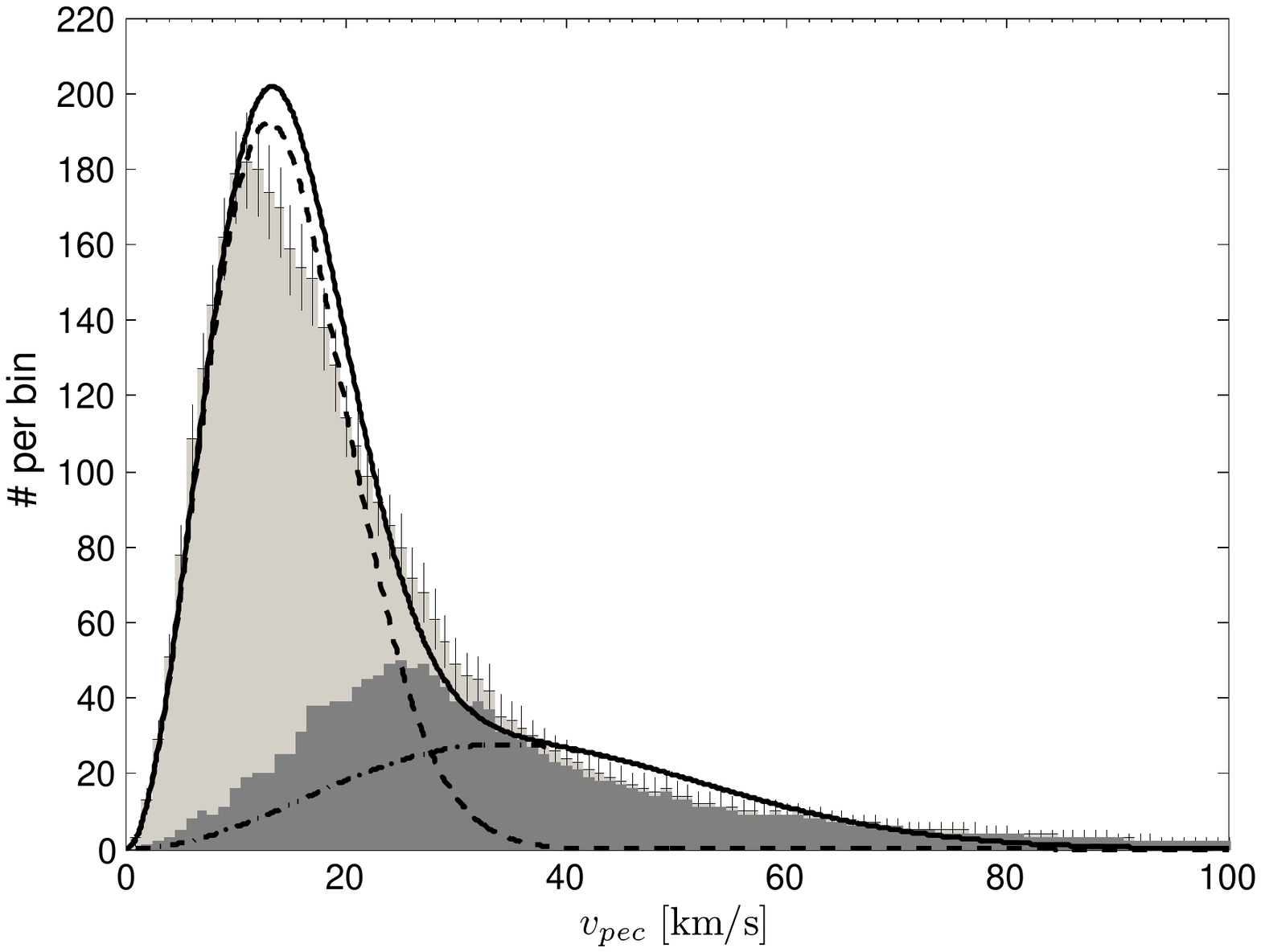}\label{subfig:vges_mitrun}}
\caption{Distributions of $v_{t,pec}$ and $v_{pec}$. Light grey histograms represent the whole star sample whereas dark grey histograms show the distributions of runaway star candidates. The velocity distributions of the high velocity group is drawn with dashed-dotted lines (cf. \autorefs{fig:2Maxwell_vtrans} and \ref{fig:2Maxwell}).}
\label{figs:hist_vges_vt_mitrun}
\end{figure}

We analysed the distributions of the peculiar velocities of 7663 young Hipparcos stars (4180 with full kinematics) in three, two as well as one dimension to identify members of the high velocity group, i.e. stars which show different kinematics than normal Population I objects and hence experienced some interaction (BSS or DES) which provides them with an additional velocity. As expected, the velocity component due to runaway formation is isotropic. \\
Performing Monte-Carlo simulations varying the observables $\pi$, $\mu_\alpha^\ast$, $\mu_\delta$ and $v_r$ within their uncertainty intervals we assigned each star a probability for being a member of the high velocity group, i.e. a runaway star. We did that for the 3D velocity $v_{pec}$, its 1D components $U$, $V$ and $W$, the 1D radial velocity $v_{r,pec}$ as well as the 2D tangential velocity $v_{t,pec}$ and its 1D components $v_{l,pec}$ and $v_{b,pec}$ to identify as many members of the high velocity group as possible, also those with a relatively low velocity. We found 2353 runaway star candidates, i.e. stars for which the runaway probability was higher than 50 per cent in at least one velocity component examined. The contamination of low velocity members is about 20 per cent at most.\\
However, those high velocity group members with very small velocities (as inferred for a Maxwellian distribution they exist) could still not be identified due to the velocity thresholds set. Therefore, we compared the velocity vector (in 3D, $\vec{v}_{pec}$, and 2D, $\vec{v}_{t,pec}$) of each individual star with the one defined by the stars in its neighbourhood or its surrounding OB association/ cluster ($\vec{v}_{neigh}$). If the velocity of the star clearly pointed away from $\vec{v}_{neigh}$ (see \autoref{fig:Skizze_Kugel}), the star was assigned being a runaway star. 184 new identifications were made.\\
Additionally, we find six additional young stars which are situated well outside any OB association/ cluster and the Galactic plane.\\
Finally, we compared our list of runaway star candidates with previously published lists and added four stars (see discussion in \autoref{appsec:otherauthors}).\\
hat gives us a total of 2547 runaway star candidates (with a contamination of low velocity group members, i.e. normal Population I stars, of 20 per cent at most). Thus the runaway frequency among young stars is approximately 27 per cent, in agreement with our theoretical expectations.\\
\Autoref{figs:hist_vges_vt_mitrun} shows the distribution of the peculiar space velocity $v_{pec}$ and the peculiar tangential velocity $v_{t,pec}$ (as \autorefs{fig:2Maxwell} and \ref{fig:2Maxwell_vtrans}) with the subsample of runaway candidates in dark grey. 
The number of runaway star candidates is somewhat higher than the number of stars belonging to the high velocity group, especially in the range of medium velocities where the two curves intersect, as to be expected. However, with this somehow conservative selection, we can be sure that we did not miss actual runaway stars. Using our combined selection we also identified runaway star candidates with relatively low velocities which certainly exist but would have not been identified by investigating only one velocity component or the absolute velocity.

We will provide \autorefs{tab:HIPid}, \ref{tab:runprob}, %\ref{tab:runcand2}, 
\ref{tab:runcand3} and \ref{tab:Zcrit} within an electronic catalogue available via VizieR.

\section*{Acknowledgments}

We would like to thank C. Dettbarn, B. Fuchs and H. Jahrei{\ss} for helping with kinematical data and J. Hern{\'a}ndez and F. Palla for providing us evolutionary models as well as M. Ammler-von Eiff for his assistance in this regard.\\
We also thank the referee, Philip Dufton, for valuable comments.\\
NT acknowledges financial support from the German National Science Foundation (Deutsche Forschungsgemeinschaft,
DFG) in grant SCHR 665/7-1 and Carl-Zeiss-Stiftung.
We acknowledge partial support from DFG in the SFB/TR-7 Gravitational
Wave Astronomy.\\
Our work has made extensive use of the Simbad and VizieR services (http://cds.u-strasbg.fr). 

\bibliographystyle{mn2e}
\bibliography{bib_runawaycat}

\appendix

\section{The velocity dispersion of the high velocity group of stars}\label{appsec:adddisp}

For normal Population I stars, i.e. members of the low velocity group, the velocity dispersion in the $z$ direction, $\sigma_W$, is smaller than the velocity dispersion in the $x$ and $y$ directions (in the Galactic plane) due to the Galactic potential that attracts the stars onto the Galactic plane. Since all stars initially belong to the low velocity group there must be a difference between $\sigma_{U/V}$ and $\sigma_W$ also for high velocity group members. Since the velocity distribution in each direction is Gaussian, 
\begin{eqnarray}
 \sigma_{H,U}^2 & = & \sigma_{L,U}^2+\sigma_{x}^2,\notag\\
 \sigma_{H,V}^2 & = & \sigma_{L,V}^2+\sigma_{x}^2,\\
 \sigma_{H,W}^2 & = & \sigma_{L,W}^2+\sigma_{x}^2,\notag
\end{eqnarray}
where $\sigma_x$ is the velocity dispersion due to runaway formation. Initially, we assume the additional velocity to be isotropic.
For the low velocity group we find that
\begin{eqnarray}
 \sigma_{L,W} & \approx & \sigma_{L,U}-\left(\unit[5]{km/s}\right)\notag\\
 \mbox{and} & &\\
 \sigma_{L,V}  & \approx & \sigma_{L,U}.\notag
\end{eqnarray}
Thus,
\begin{equation}
 \sigma_{H,W}^2 = \sigma_{H,U/V}^2-\unit[10]{km/s}\cdot\sigma_{L,W}-\unit[25]{km^2/s^2}.
\end{equation}
With $\sigma_{H,U/V}\approx\unit[25]{km/s}$ and $\sigma_{L,W}\approx\unit[5]{km/s}$, it follows that $\sigma_{H,W}\approx\unit[23]{km/s}$ theoretically. However, since the runaway formation occurred some time in the past (for BSS runaways in the sample this timespan might be comparable with the age of the star before the supernova), the Galactic potential makes an impact on the higher velocities. For that reason, we expect to measure a somewhat lower value $\sigma_{W,H}$ for the high velocity group than the predicted value of $\unit[23]{km/s}$, thus the found value of $\sigma_{W,H}\approx\unit[17]{km/s}$ is in good agreement with our predictions and we conclude that runaway formation leads to an additional velocity that is isotropic.

\section{An individual discussion on runaway stars found in the literature}\label{appsec:otherauthors}

The classical runaway HIP 102195 was not identified by us since its peculiar spatial velocity is rather small ($v_{pec}=\unit[16{.}6^{+2{.}9}_{-7{.}1}]{km/s}$). Note that \citet{1961BAN....15..265B} also quote a small velocity ($\approx\unit[23]{km/s}$). As proposed by the authors, HIP 102195 apparently originated from the Lacerta OB1 association. However, using 3D data, we cannot confirm this origin but instead find that in $\unit[10{.}0]{\%}$ of 10000 Monte-Carlo runs the star's position is located within the boundaries of Cygnus OB7 about $\unit[11{.}1^{+1{.}9}_{-3{.}1}]{Myr}$ in the past which is in excellent agreement with the association age of $\unit[13]{Myr}$ \citep{2001A&A...371..675U}. Due to the large number of parameters involved (position and velocity of the star and association) the fraction of successful runs is expected to be small \citep[cf.][]{2001AA...365...49H,2010MNRAS.402.2369T}. Hence, we include this star into our runaway star sample.\\
The twelve stars identified by \citet{1974RMxAA...1..211C}, \citet{1979ApJ...232..520S} and \citet{2005AA...431L...1M} are not re-identified by us simply because the authors used photometric distances which are systematically too large, thus generating large peculiar velocities (this can be directly seen from comparison between Columns 5 and 6 of Table 1 in \citealt{2005AA...431L...1M}), whereas we used parallactic distances to determine peculiar velocities. \\
HIP 67279 was included by \citet{1990AJ.....99..608L} owing to its large distance from the Galactic plane of $z=\unit[1]{kpc}$ (according to the author's definition of a runaway star $z$ must be larger than $\unit[20]{km/s}$ times the main-sequence lifetime of the star, cf. \autoref{footn:Zcrit}). The photometric distance of $\unit[1{.}17]{kpc}$ that the authors adopted from \citet{1975MNRAS.171..353K} is however once more too large and the actual distance from the Galactic plane derived from the parallax (parallactic distance is $\unit[472^{+182}_{-103}]{pc}$) is $z=\unit[397^{+80}_{-130}]{pc}$. With this $z$ and an age $\tau_\star=\unit[0{.}3\pm2{.}1]{Myr}$ as inferred from evolutionary models (see \autoref{sec:sample}), HIP 67279 would need a vertical velocity component as large as $W=\unit[1300]{km/s}$ to have originated from the Galactic plane. Similarly, \citet{2005AA...437..247D} identified the three stars listed in \autoref{tab:missingrun} only from their distance to the Galactic plane again using photometric distances. With the better parallactic distances they do not satisfy the criterion applied by the authors ($z>\unit[250]{pc}$).\\
Four of the seven runaway candidates listed by \citet{2001AA...365...49H} (HIP 20330, 86768, 92609, 103206) are not recognised as runaway stars by us due to different input data (especially $\pi$) as
\citet{2001AA...365...49H} used the old Hipparcos reduction \citep{1997A&A...323L..49P} (smaller $\pi$ in all four cases) whereas we used the latest published data by \citet{2007AA...474..653V}. In the case of HIP 48943, the radial velocity adopted by \citet{2001AA...365...49H} of $v_r=\unit[39{.}0\pm5{.}0]{km/s}$ differs from our value of $v_r=\unit[29{.}6\pm3{.}6]{km/s}$ \citep{2007AN....328..889K} resulting in different peculiar space velocities ($\unit[30{.}7^{+4{.}9}_{-5{.}1}]{km/s}$ and $\unit[22{.}6^{+3{.}9}_{-4{.}1}]{km/s}$, respectively). Since HIP 48943 is an astrometric binary \citep{2007ApJS..169..105M}, we account for uncertainties in the radial velocity and include it into our sample of runaway star candidates. For another one, HIP 38455, \citet{2001AA...365...49H} adopted $v_r=\unit[-31{.}0\pm5{.}0]{km/s}$. According to \citet{1986Ap&SS.121..205H} HIP 38455 is a spectroscopic binary with a systemic radial velocity of $\unit[29{.}5]{km/s}$. This significantly different radial velocity changes the peculiar spatial velocity dramatically ($v_r=\unit[-31]{km/s}$: $v_{pec}=\unit[47]{km/s}$, $v_r=\unit[29{.}5]{km/s}$: $v_{pec}=\unit[14]{km/s}$), hence the star is not a runaway. Moreover, \citet{2001AA...365...49H} corrected the velocities for Solar motion using the LSR published by \citet{1998MNRAS.298..387D} which results in somewhat larger space velocities than ours (which is why we did not find some of their runaway stars) but does not accurately reflect the motion of young stars relative to the Sun (see \autoref{sec:sample}). For HIP 3881 \citet{2001AA...365...49H} suggested a birth association (Lacerta OB1). We find that in 1{.}8 per cent of 10000 Monte-Carlo runs the star's position coincided with the boundaries of Lacerta OB1 ($\unit[6{.}3^{+0{.}8}_{-1{.}2}]{Myr}$ in the past). Like HIP 102195 (see above), we include this star into our runaway star sample. In addition, we include HIP 26241 (= $\iota$ Ori, highly eccentric spectroscopic binary) since it was very probably part of a former triple system (together with the classical runaways HIP 24575 = AE Aur and HIP 27204 = $\mu$ Col), thus ejected via DSS from the Trapezium cluster \citep[e.g.][]{2001AA...365...49H} and a member of the high velocity group.

\label{lastpage}

\end{document}